\documentclass[draftcls,letterpaper,12pt,onecolumn]{IEEEtran}

\usepackage{amsmath}
\usepackage{amsfonts,amssymb}
\usepackage{epsfig}
\usepackage{subfigure}
\usepackage{cite}

\def\BibTeX{{\rmfamily B\kern-.05em{\scshape i\kern-.025em
b}\kern-.08em \TeX}}

\newtheorem{theorem}{Theorem}
\newtheorem{lemma}{Lemma}

\newcommand{\e}{\epsilon}

\newtheorem{proposition}{Proposition}

\newcommand{\bz}{\mathbf{z}}
\newcommand{\bs}{\mathbf{S}}
\newcommand{\by}{\mathbf{y}}

\newcommand{\bI}{\mathbf{I}}

\newcommand{\mhat}{\hat{\mu}}
\newcommand{\sh}{\sigma_h^2}
\newcommand{\sz}{\sigma_z^2}
\newcommand{\emax}{\epsilon_{max}}
\IEEEoverridecommandlockouts

\begin{document}
\title{Adaptive Training for Correlated Fading Channels with 
Feedback\thanks{This work was supported
by the U.S. Army Research Office under grant
W911NF-07-1-0028 and NSF under grant CCR-0310809. 
This paper was
presented in part at the IEEE International
Symposium on Information Theory (ISIT), Seattle, USA, July, 2006 and at the
Information Theory and its Applications (ITA), San Diego, January, 2008.}}

\author{\authorblockN{Manish Agarwal, Michael Honig, and Baris Ata\\}
\authorblockA{
Northwestern University\\
\{m-agarwal,mh,b-ata\}@northwestern.edu }\\
\today }
\maketitle

\vspace{-0.7in}

\begin{abstract}
We consider data transmission through 
a time-selective, correlated (first-order Markov) 
Rayleigh fading channel subject to an average power constraint.
The channel is estimated at the receiver with a pilot signal, and the 
estimate is fed back to the transmitter.  The estimate is used for 
coherent demodulation, and to adapt the data and pilot powers. 
We explicitly determine the optimal pilot and data power control policies
in a continuous-time limit where the channel state
evolves as an Ornstein-Uhlenbeck diffusion process,
and is estimated by a Kalman filter at the receiver.
The optimal pilot policy switches between zero and 
the maximum (peak-constrained) value (``bang-bang'' control), 
and approximates the optimal discrete-time policy at 
low Signal-to-Noise Ratios (equivalently, large bandwidths).
The switching boundary is defined in terms of
the system state (estimated channel mean and associated error variance),
and can be explicitly computed.
Under the optimal policy, the transmitter conserves power 
by decreasing the training power when the
channel is faded, thereby increasing the data rate.
Numerical results show a significant 
increase in achievable rate due to the adaptive training scheme 
with feedback, relative to constant (non-adaptive) training,
which does not require feedback.
The gain is more pronounced at relatively low SNRs and with fast
fading. Results are further verified through Monte Carlo simulations.
\end{abstract}


\section{Introduction}
\label{intro}

The achievable rate for a time-selective fading channel
depends on what channel state information (CSI) is available
at the receiver and transmitter. Namely, CSI at the receiver
can increase the rate by allowing coherent detection,
and CSI at the transmitter allows adaptive rate
and power control (e.g., see \cite[Ch. 6]{TseVis05}).
Obtaining CSI at the receiver and/or transmitter
requires overhead in the form of a pilot signal and feedback.

We consider a correlated time-selective flat Rayleigh fading
channel, which is unknown at both the receiver and transmitter.
The transmitter divides its power between a pilot,
used to estimate the channel at the receiver, and the data.
Given an average transmitted power constraint,
our problem is to optimize the instantaneous pilot and data powers
as functions of the time-varying channel realization.
Our performance objective is a lower bound on the achievable rate,
which accounts for the channel estimation error.

Power control with channel state feedback, assuming
the channel is perfectly known 
at the receiver, has been considered in 
\cite{GolVar97IT,CaiSha99IT,Viswan99IT,LiuEli04ACC,CaiTar99IT}.
There the focus is on optimizing the input
distribution for different channel models using
criteria such as rate maximization and outage
minimization. Optimal power allocation
in the presence of channel estimation error
has been considered in \cite{KleGal01ISIT, YooGol06IT}.
The problem of optimal pilot design for a variety of fading channel 
models has been considered in \cite{Schram98TCOM,
Cavers91TVT,Medard00IT,OhnGia02TWC,HasHoc03IT,
BdeAbo04ICC,DonTon04TSP}. There the pilot power
and placement, once optimized, is fixed and is not adapted with
the channel conditions.
A key difference here is that the transmitter 
uses the CSI to adapt \emph{jointly} the 
instantaneous data and pilot powers.
Because the channel is correlated in time,
adapting the pilot power with the estimated channel state
can increase the achievable rate.
We also remark that although we analyze a single narrowband
fading channel, our results apply to a set
of parallel fading Gaussian channels, where the 
average power is split over all channels.

We start with a correlated block fading model in which the
sequence of channel gains is Gauss-Markov\footnote{Several 
theoretical and measurement based studies, such as 
\cite{WanMoa95TVT,WanCha96TVT, TanBea00TCOM, ZhaKas99TCOM},
have argued that this is a
reasonable model for time-selective wireless channels.}
with known statistics at the receiver.
The channel estimate is updated at the beginning of each block
using a Kalman filter, and determines
the power for the data, and the power for the pilot symbols
in the succeeding coherence block.
Optimal power control policies are specified
implicitly through a Bellman equation \cite{Bertse95}.
Other dynamic programming formulations of
power control problems have been presented in
\cite{CaiTar99IT,NegCio02IT,ZafMod05Allerton,ChaDjo05IT, BerGal02IT},
although in that work the channel is either known perfectly 
(perhaps with a delay), or is unknown and not estimated.

Because an analytical solution to the Bellman equation
appears to be difficult to obtain, we study
a diffusion limit in which 
the correlation between successive coherence
blocks tends to one and the average power goes to zero.
(This corresponds to a wideband channel model in
which the available power is divided uniformly over a large
number of parallel flat Rayleigh fading sub-channels.)
In this limit, the
Gauss-Markov channel becomes
a continuous-time Ornstein-Uhlenbeck process \cite{Oksend03},
and the Bellman equation becomes a partial differential equation (PDE).
A diffusion equation is also derived,
which describes the evolution of the state
(channel estimate and the associated error variance),
given a power allocation policy.
In this limit, we show that given a peak power
constraint for the pilot power, the optimal pilot power control policy
is a switching policy (``bang-bang'' control): the pilot power is either
the maximum allowable or zero,
depending upon the current state.
Hence the optimal pilot power control policy
requires at most one feedback bit per coherence block.
Also, the optimal data power control policy
is found to be a variation of waterfilling \cite{TseVis05}.
Other work in which the wireless channel is modeled
as a diffusion process is presented in \cite{FenFie07TCOM,ChaDjo05IT}.

The switching points for the optimal policy
form a contour in the state space, which
is referred to as the \emph{free boundary}
for the corresponding PDE. Solving this PDE then
falls in the class of {\em free boundary problems} \cite{Zwilli98, Friedm82}.
We show that in the diffusion limit the system state 
becomes confined to a narrow region along the boundary.
Furthermore, the associated probability distribution
over the boundary is exponential. That enables a numerical
characterization of the boundary shape.

Our results show that the average pilot power should decrease
as the channel becomes more severely faded.
We observe that the optimal switching policy
is equivalent to adapting the pilot symbol insertion rate 
with {\em fixed} pilot symbol energy.\footnote{Alternatively,
the same performance can be achieved by fixing the pilot insertion rate 
and varying the pilot power. However, in principle that would require
infinite-precision feedback in contrast to bang-bang control, 
which requires one feedback bit per coherence block.}
The optimal pilot insertion rate as a function of the
channel estimate is then determined by the 
shape of the free boundary. We show
that the boundary shape essentially shifts pilot power 
from more probable (faded) states to less probable (good) states.
Furthermore, the boundary shape guarantees that the channel estimate 
is sufficiently accurate to guide the power adaptation.  

Numerical results show that pilot power adaptation
can provide substantial gains 
in achievable rates (up to a factor of two).
The gains are more pronounced at low SNRs and with fast fading channels.
Although these results are derived in the limit of large bandwidth (low SNR),
Monte Carlo simulations show that they provide an accurate estimate of
the performance when the bandwidth is large but finite 
(a few hundred coherence bands).  
Moreover, the optimal switching policy in the diffusion limit
accurately approximates the optimal pilot power control policy
for the discrete-time model, and provides essentially the same
performance gains relative to constant pilot power.

To limit the overall feedback rate, we also consider combining
the adaptive pilot power with ``on-off'' data power control,
which also switches between a fixed positive value and zero.
(Hence that also requires at most one bit feedback per coherence block.)
The corresponding optimal free boundaries are computed,
and results show that this scheme gives negligible loss 
in the achievable rate. 


%

%

The next section presents the system model and Section \ref{discrete}
formulates the pilot optimization problem as a dynamic program.
Section \ref{cont} presents the associated diffusion limit and the
corresponding Bellman equation.
The optimal policy is then characterized in 
Sections \ref{solanal}-\ref{water} with optimal data power
control, and in Section \ref{on-off} with optimal 
on-off data power control. Numerical results showing free boundaries
and the corresponding performance are also presented in Sections \ref{water}
and \ref{on-off}. Training overhead is discussed in Section \ref{tl},
and conclusions and remaining issues are discussed in Section \ref{conc}.

\section{Correlated Block Fading Model}
\label{model}

We start with a block fading channel model in which
each coherence block contains $M$ symbols,
consisting of $T$ pilot symbols and $D$ data symbols.
The vector of channel outputs for coherence block
$i$ is given by
\begin{equation}
\by_i = h_i \left(
\begin{array}{c}
\sqrt{P_{i;T}} \bs_{i;T} \\
\sqrt{P_i} \bs_i
\end{array} \right)
+ \bz_i
\label{eq:model}
\end{equation}
where $\bs_{i;T}$ and $\bs_i$ are, respectively,
vectors containing the pilot and data symbols,
each with unit variance,
and $P_{i;T}$ and $P_i$ are the associated pilot and data powers.
The noise $\bz_i$ contains
circularly symmetric complex Gaussian (CSCG) random variables,
and is white with covariance $\sigma_z^2 \bI$.
The channel gain $h_i$ is also CSCG, is constant within the block,
and evolves from block to block according to a Gauss-Markov
process, i.e.,
\begin{equation}\label{eq:gm}
h_{i+1} = r\,h_i + \sqrt{1 - r^2} \,\, w_i
\end{equation}
where $w_i$ is an independent CSCG random variable
with mean zero and variance ${\sigma_h}^2$, and
$r \in [0,1]$ determines the correlation between successive blocks.
We will assume that $r$ and $\sigma_h^2$ are known at
the receiver. 
The training energy per symbol in block $i$ is defined as
$\epsilon_i = \alpha P_{i;T}$, where $\alpha = T/M$.
In what follows, it will be convenient to
write $P_{i;T}$ as $\epsilon_i / \alpha$.


The receiver updates the channel estimate during
each coherence block with a Kalman filter \cite{BroHwa97},
given the model (\ref{eq:gm}) and the pilot symbols, 
and relays the estimate back to the transmitter.
The feedback occurs between the pilot and data symbols,
and is assumed to occupy an insignificant fraction
of the coherence time. 
We re-write the noise vector $\bz_i$ as 
$[\bz_{i;T}^\dag \,\, \bz_{i}^\dag]^\dag$ where $\bz_{i;T}$ and $\bz_{i}$
are, respectively, $T \times 1$ and $D \times 1$ vectors
and $[\cdot]^\dag$ denotes Hermitian transpose. 
The channel estimate $\hat{h}_{i}$ and estimation error
$\theta_{i} = E(|{h_{i}}|^2) - E(|\hat{h}_{i}|^2)$
evolve according to the following Kalman filter updates:
\begin{eqnarray}\label{eq:kalest}
\hat{h}_{i+1} & = & r  \hat{h}_i +
g_{i+1}  \sqrt{\frac{\epsilon_{i+1}}{\alpha}}   T   e_{i+1|i}
+ g_{i+1} \bs_{i+1;T}^\dag    \bz_{i+1;T} \\
\label{eq:kalvar}
\theta_{i+1} & = & \frac{{\sigma_z}^2 \,
\theta_{i+1|i}}{\epsilon_i\,M\,\theta_{i+1|i} + {\sigma_z}^2}
\end{eqnarray}
where
\begin{eqnarray}\label{eq:kalgain}
g_{i} & = &
\sqrt{\frac{\epsilon_i}{\alpha}}\, \frac{\theta_{i}}{{\sigma_z}^2}\\
\label{eq:cerr}
e_{i+1|i} & = & h_{i+1} - r\,\hat{h}_i \\
\label{eq:condvar}
\theta_{i+1|i} & = & r^2\,\theta_i + (1 - r^2)\,{\sigma_h}^2 .
\end{eqnarray}
It is straightforward to show that 
the channel estimate $\hat{h}_i$ in (\ref{eq:kalest}) 
does not depend on $T$. (What is important is the total
pilot energy per coherence block.)
Hence the data rate is maximized by taking $T =1$
with fixed $\epsilon_i$ (i.e., the training power $P_{i;T}  = \epsilon_i M$).
Training therefore requires an overhead of 
$1/M$ fraction of the channel uses. We ignore this overhead
for the time being and focus on optimizing the training power
$\epsilon_i$. This issue is revisited in Sec. \ref{tl}.


We wish to determine $P_i$ and $\epsilon_i$, which maximize
the achievable rate. Specifically, the channel estimate $\hat{h}_i$
and variance $\theta_i$ determine the data power in the
{\em current} coherence block, $P_i$,
and the pilot power in the {\em next} coherence block, $\epsilon_{i+1}$.
We assume that the transmitter codes over many coherence blocks,
and use the following lower bound on ergodic capacity,
which accounts for channel estimation error, as the 
performance objective \cite{Medard00IT,AgaHon05Allerton}, 
\begin{equation}\label{eq:minfo}
R(P_i, \hat{\mu}_i, \theta_i) =
\log \left( 1 + \frac{P_i\, \hat{\mu}_i}{P_i\,\theta_i +
{\sigma_z}^2} \right)
\end{equation}
where $\hat{\mu}_i = |\hat{h}_i|^2$. In the next section, we
formulate the joint pilot and data power optimization problem,
and subsequently characterize the optimal power control policy
implicitly as the solution to a discrete-time Bellman equation.

\section{Dynamic Programming Formulation}
\label{discrete}
The pilot and data power control problem can be stated as
\begin{eqnarray}
\begin{aligned}
\label{eq:dtmc}
\max_{\{P_i, {\epsilon}_i \}}\, & \liminf_{n \rightarrow \infty} \,\,
\frac{1}{n} E\left[ \sum_{i=0}^{n-1} R(P_i, \hat{\mu}_i, \theta_i)\right] \\
\text{subject to:} & \,\,\label{eq:pconst} \limsup_{n \rightarrow
\infty}E\left[ \frac{1}{n} \sum_{i=0}^{n-1} \,{\epsilon}_i +
\frac{1}{n}\sum_{i=0}^{n-1} \,P_i \right] \le \,P_{av},\\
\text{and} & \qquad\epsilon_i \le \emax,
\end{aligned}
\end{eqnarray}
where the expectation is over the sequence of channel gains.
We have imposed an additional peak power
constraint on the training power.
This is a discrete-time Markov control problem,
so that the solution can be formulated as an infinite-horizon 
dynamic program with an average value objective.
The system state at time (block) $i$ is
$S_i = ({\hat{\mu}}_i, \theta_i)$,
and the action maps the state to
the power pair $(P_i, {\epsilon}_{i+1})$.
To see that $S_i$ is the system state,
note that $e_{i+1|i}$ in \eqref{eq:cerr} and $\hat{h}_i$
are independent random variables, hence it follows from
\eqref{eq:kalest} and \eqref{eq:kalvar} that
the probability distribution of $S_{i+1}$
is determined only by $S_i$ and the action ${\epsilon}_{i+1}$.
The process $\{({\hat{\mu}}_i, \theta_i)\}$ is therefore
a Markov chain driven by the control $\{{\epsilon}_i\}$.

The average power constraint in \eqref{eq:pconst}
can be included in the objective through a Lagrange
multiplier giving the relaxed problem
\begin{equation}\label{eq:lag}
\max_{P_i,\, 0\le{\epsilon}_i\le\emax}\,
\liminf_{n \rightarrow \infty}\,\,
\frac{1}{n} E\left[ \sum_{i=0}^{n-1}\,
\left[R(P_i, \hat{\mu}_i, \theta_i) - \lambda \,({\epsilon}_i + P_i )
\right]\right]
\end{equation}
where $\lambda$ is chosen to enforce the constraint \eqref{eq:pconst}.
If there exists a bounded function $V(\hat{\mu}, \theta)$ and
a constant $C$, which satify the Bellman equation
\begin{equation}\label{eq:bell1}
V(\hat{\mu}, \theta) + C =
\max_{P,\,0\le {\epsilon}\le \emax}\,\left[R(P, \hat{\mu}, \theta)
- \lambda \,({\epsilon} + P ) + E_{\epsilon, (\hat{\mu}, \theta)} [V] \right] ,
\end{equation}
then an optimal policy maximizes the right-hand side \cite{Bertse95}.
The function $V(\cdot,\cdot)$ is called an ``auxiliary value function'',
and $C$ is the maximum value of the objective in \eqref{eq:lag}.
The expectation $ E_{\epsilon, (\hat{\mu}, \theta)} [\cdot]$
is over the conditional probability of
$S_{i+1}$ given $S_i = (\hat{\mu}, \theta)$
and action $\epsilon_{i+1}=\epsilon$.

Using the channel state evolution equations derived
in Section\ \ref{model}, we have
\begin{equation}\label{eq:expnxt}
E_{\epsilon, (\hat{\mu}, \theta)} [V]
= \int_0^{\infty}\, V(u, \theta_{i+1})\, f_{\mhat_{i+1} | S_i} (u) du
\end{equation}
where $f_{\mhat_{i+1} | S_i} (u)$ is the conditional density of
$\hat{\mu}_{i+1} = |\hat{h}_{i+1}|^2$ given
$S_i = (\hat{\mu}_i, \theta_i) = (\hat{\mu}, \theta)$,
and $\theta_{i+1}$ is given by
\eqref{eq:kalvar}.
From \eqref{eq:kalest} it follows that
$f_{\mhat_{i+1} | S_i} (u)$ is Ricean with noncentrality parameter $r^2 \hat{\mu}$ and variance
$[(\theta_{i+1|i} \epsilon M)/\sigma_z^2] \theta_{i+1} + r^2 \hat{\mu}$,
where $\theta_{i+1}$ and $\theta_{i+1|i}$ are
given by \eqref{eq:kalvar} and \eqref{eq:condvar}, respectively.

\section{Diffusion Limit}
\label{cont}
The Bellman equation \eqref{eq:bell1} is an
integral fixed point equation, and appears to be
difficult to solve analytically. To gain insight
into properties of optimal policies, we consider 
the following scaling, corresponding to a low SNR regime:
\begin{enumerate} 
\item Time is scaled by the factor $1/N$, where
$N \gg M$ is large, so that each coherence block of $M$ symbols
corresponds to $\delta t = \frac{M}{N}$ time units.
Therefore one time unit in the scaled system contains
$\frac{1}{\delta t} = \frac{N}{M}$ coherence blocks, 
or equivalently $N$ channels uses.
\item The correlation between adjacent coherence blocks
is $r = 1 - \rho (\delta t)$, where $\rho$ is a constant. 
Hence this correlation goes to one as $N \to \infty$
(equivalently, the channel coherence time goes to zero),
but with fixed correlation between blocks 
separated by $N$ channel uses.
\item To maintain constant energy over $N$ channel uses,
the average power $P_{av}$, data power $P_i$, training 
power $\epsilon_i$ and the maximum training 
power $\epsilon_{max}$ are each scaled by $1/N$. 
\end{enumerate}


In the limit as $N \to \infty$, it can be shown that
the discrete-time, complex, Gauss-Markov process $\{h_i\}$,
given by \eqref{eq:gm},
converges weakly to a continuous-time Ornstein-Uhlenbeck
diffusion process $h(t)$ (e.g., see \cite[Ch. 8]{Whitt02}).
Furthermore, the limiting channel process satisfies
the stochastic differential equation (SDE)
\begin{equation}\label{eq:ou}
d\,h(t) = -\rho\,h(t)\,dt + \sqrt{2 \rho} \sigma_h \,dB(t),
\end{equation}
where $B(t)$ is complex Brownian motion, and we assume that
the initial state $h(0)$ is a CSCG random variable
with zero mean and variance $\sigma_h^2$. 
This is a Gauss-Markov process, which is
continuous in probability, and has autocorrelation function
\begin{equation}\label{eq:corr}
\Phi (\tau) = {\sigma_h}^2\, e^{-\rho\tau},
\end{equation}
where $\tau$ is the lag between the time samples of the channel.  
Hence $\rho$ determines how fast the channel varies
relative to the symbol rate. For example, the end-to-end
normalized correlation across an interval of $\tau =1$ (or, equivalently
$N$ channel uses) is $e^{-\rho}$. 

The diffusion limit considered can be interpreted
as \emph{zooming out} on the channel and associated data transmission.
Segments of the discrete channel process then become ``compressed'' 
in time, but with increasing correlation between successive
coherence blocks so that the channel autocorrelation remains fixed.
Prior work, which advocates the use of diffusion
models for wireless channels is presented 
in \cite{FenFie07TCOM,ChaDjo05IT}.
In this limit the Kalman filter continuously
estimates the channel process, and the pilot and
data powers are continuously updated based on continuous feedback.
(We will see that to achieve optimal performance
the feedback need not be continuous.)
The optimal power control policy in the diffusion limit
can then be interpreted as an approximation for the optimal
discrete-time policy. (This will be illustrated numerically.)

The power scaling by $1/N$ can be 
interpreted as introducing $N$ parallel,
independent and statistically identical \emph{sub-channels}  
over which the power is equally split. 
Hence the low SNR regime corresponds to a {\em wideband} 
channel.\footnote{This bandwidth scaling is simply an 
interpretation of the effect of the power scaling. 
The diffusion process is still associated with 
a {\em flat} Rayleigh fading channel.}

In the diffusion limit the channel estimate and estimation error updates
given by \eqref{eq:kalest} and \eqref{eq:kalvar}, respectively,
become the dynamic equations
\begin{eqnarray}\label{eq:chan}
d\hat{h}(t) & = & - \rho\,\hat{h}(t)\, dt +
\theta (t) \sqrt{\frac{\epsilon (t)}{{\sigma_z}^2}} \,\, d \overline{B}(t), \\
\label{eq:errdyn}
\frac{d\theta (t)}{dt} & = &
 2\rho \,(\sigma_h^2- \theta (t)) -
\frac{\epsilon (t) {\theta}^2 (t)}{{\sigma_z}^2},
\end{eqnarray}
where $\overline{B}(t)$ is a complex Brownian motion
independent of $B(t)$, and
$\epsilon (t)$ is the pilot power at time $t$.
A heuristic derivation of \eqref{eq:ou}, \eqref{eq:chan} and \eqref{eq:errdyn}
from the discrete-time equations \eqref{eq:gm}, \eqref{eq:kalest}
and \eqref{eq:kalvar} is given in Appendix \ref{ap:lmt}. 

Note that both $h(t) = h_r(t) + j\, h_j(t) $ and 
$\hat{h}(t)  = \hat{h}_r(t) + j\, \hat{h}_j(t) $ are complex.
The following SDE defining the evolution of the channel estimate
$\hat{\mu}(t) = |\hat{h}(t)|^2= \hat{h}_r^2(t) + \hat{h}_j^2(t)$ can 
be obtained from \eqref{eq:chan} and a straightforward
application of Ito's Lemma \cite[Ch. 4]{Oksend03}:
\begin{equation}\label{eq:estevol0}
d\hat{\mu}(t) = \left[- 2 \rho\,\hat{\mu}(t) + 
\frac{\epsilon (t)\theta^2(t)}{{\sigma_z}^2} \right]\,dt 
 + 2 \left[\theta (t) \sqrt{\frac{\epsilon (t)}{{\sigma_z}^2}} 
\hat{h}_r(t)\right] d\overline{B}_{r}(t) + 
2 \left[\theta (t) \sqrt{\frac{\epsilon (t)}{{\sigma_z}^2}} 
\hat{h}_j(t)\right] d\overline{B}_{j}(t),
\end{equation}
where $\overline{B}_{r}(t)$ and $\overline{B}_{j}(t)$ 
are, respectively, the real and imaginary parts of the
complex Brownian Motion $\overline{B}(t)$. 
We can re-write \eqref{eq:estevol0} as
\begin{equation}\label{eq:estevol1}
d\hat{\mu}(t) = \left[- 2 \rho\,\hat{\mu}(t) + 
\frac{\epsilon (t)\theta^2(t)}{{\sigma_z}^2} \right]\,dt 
 + \theta (t) \sqrt{\frac{\epsilon (t) \mhat(t)}{{\sigma_z}^2}} \,\,  
d\underline{B}(t),
\end{equation}
where $d\underline{B}(t)$ is a \emph{real}-valued 
standard Brownian motion.

\begin{lemma}\label{lm:cont}
Given $\e(t) \in [0, \emax]$, the state process 
$S(t) = (\mhat(t), \theta(t))$,
which is the solution to the stochastic differential equations 
\eqref{eq:estevol1} and \eqref{eq:errdyn}, has continuous sample paths.
\end{lemma}

The proof is given in Appendix \ref{ap:cont}.
Note that this Lemma does not require the control input
$\e(t)$ to be continuous in time. This observation will be
useful in the subsequent discussion. 

We now consider the continuous-time limit of the 
optimization problem \eqref{eq:dtmc}.
If the data power for the $i^{th}$ discrete coherence block 
is $P_i/N$, then for large (but finite) $N$, 
the objective $R(P_i/N , \hat{\mu}_i,\theta_i)$
becomes close to the continuous objective
$R[ P(t)/N, \hat{\mu}(t), \theta(t)]$, where $R(\cdot)$
is given by \eqref{eq:minfo} and the index $i$
corresponds to time $t$. 
Our problem is then to choose $\epsilon(t)$ and $P(t)$, as a
function of the state $(\hat{\mu}(t), \theta(t))$,
to maximize the accumulated rate function (over time),
averaged over the channel process $h(t)$. 
Equivalently, we can maximize the scaled objective
$N\,R[ P(t)/N, \hat{\mu}(t), \theta(t)]$ 
(corresponding to the sum rate over $N$ parallel sub-channels).
A difficulty is that this objective is unbounded
as $N\to\infty$.
To simplify the analysis, we first take $N =1$,
which lower bounds the objective 
for all $N \ge 1$ (and corresponds to scaling
up the power in the diffusion limit).
After characterizing the optimal policy
we then replace the objective with the preceding
scaled objective with fixed $N$ to generate 
numerical results.\footnote{In Sec. \ref{onoff:analytic}
we discuss the rate of growth of the scaled rate
objective as $N\to\infty$.}

We therefore rewrite the discrete-time optimization \eqref{eq:dtmc} as
the continous-time control problem
\begin{eqnarray}
\begin{aligned}
\label{eq:diff}
\max_{(P(t), {\epsilon}(t))}\,  &
\liminf_{t \rightarrow \infty} \,\,
\frac{1}{t} E\left[\int_{0}^{t} R(P(t),
\hat{\mu}(t), \theta(t)) dt \right]  \\
\text{subject to:} &\,\, \label{eq:pconst1} \limsup_{t \rightarrow
\infty}E\left[ \frac{1}{t} \int_{0}^{t} \,{\epsilon}(t)\,dt +
\frac{1}{t}\int_{0}^{t} \,P(t)\,dt \right] \le \,P_{av},\\
\text{and}& \qquad \e(t) \le \emax  .
\end{aligned}
\end{eqnarray}
Analogous to \eqref{eq:bell1}, the Bellman equation
can be written as (see \cite{DynYus79})
\begin{equation}\label{eq:bell2}
C = \max_{P,\, 0\le{\epsilon}\le \emax}\,\left\{ R(P, \hat{\mu}, \theta) -
\lambda \,({\epsilon} + P) + A_{\epsilon}[V(\hat{\mu}, \theta)] \right\}
\end{equation}
where $A_{\epsilon}$ is the generator of the
state process $(\hat{\mu}(t), \theta (t))$
with pilot power $\epsilon(t)$ \cite[Ch. 7]{Oksend03},
and is given by
\begin{equation}\label{eq:gen}
A_{\epsilon}[V] = \frac{E[dV]}{dt} =  a + \epsilon b
\end{equation}
where
\begin{eqnarray}
a & = &  \frac{\partial V}{\partial \hat{\mu}}
\left( - 2 \rho \hat{\mu} \right) +
\frac{\partial V}{\partial \theta}
\left( - 2 \rho \theta + 2 \rho {\sigma_h}^2  \right) \\
\label{eq:diffop}
b & = &  \frac{{\theta}^2}{{\sigma_z}^2}
\left[ \frac{\partial V}{\partial \hat{\mu}} -
\frac{\partial V}{\partial \theta} +
\hat{\mu} \frac{\partial^2 V}{\partial \hat{\mu}^2} \right]
\end{eqnarray}
and the dependence on $t$ is omitted for notational convenience.
Here we ignore existence issues, and simply assume that there
exists a bounded, continuous, and
twice differentiable function $V(\cdot,\cdot)$ satisfying \eqref{eq:bell2}.
Note that $V(\cdot,\cdot)$ is 
unique only up to a constant \cite[Ch. 4]{Bertse95},\cite{DynYus79}.

\begin{theorem}
\label{thm}
Given the pilot power constraint $\epsilon \in [0, \epsilon_{max}]$,
the optimal pilot power control policy is given by
\begin{equation}\label{eq:bbc}
\epsilon^{\star} = \left \{ \begin{array}{ll}
  \epsilon_{max} & \textrm{if $ b - \lambda \ge 0$}\\
  0 & \textrm{otherwise}.
\end{array} \right.
\end{equation}
\end{theorem}

In words, optimal pilot power control is achieved
by a switching (bang-bang) policy.
This follows immediately from
substituting the generator $A_\epsilon$,
given by \eqref{eq:gen}-\eqref{eq:diffop}, into \eqref{eq:bell2}, i.e.,
\begin{equation}\label{eq:bellmid}
C = J(\hat{\mu}, \theta, \lambda) +
\max_{\epsilon} \left[ a + \epsilon (b- \lambda) \right]
\end{equation}
where
$
J(\hat{\mu}, \theta, \lambda) =
\max_{P} \left[ R(P, \hat{\mu}, \theta) - \lambda P \right]
$.
Substituting \eqref{eq:bbc} into
\eqref{eq:bellmid} gives the final version of the Bellman equation
\begin{equation}\label{eq:bellfinal1}
C = J(\hat{\mu}, \theta, \lambda) +  a + \epsilon_{max} (b- \lambda)^{+}
\end{equation}
where $(x)^+ = \max\{0,x\}$.
An alternative way to arrive at 
\eqref{eq:bellmid} and \eqref{eq:bellfinal1}
is to take the diffusion limit of the discrete-time 
Bellman equation \eqref{eq:bell1}.
This alternative derivation is given in Appendix \ref{ap:alt}.

It is easily shown that the optimal data power allocation is
\begin{eqnarray}
P_d(\mhat, \theta, \lambda) & = & \arg \max_{P}
\left[ R(P, \hat{\mu}, \theta) - \lambda P \right] \\
& = & \biggr( \frac{- \lambda {\sigma_z}^2
(2 \theta + \hat{\mu}) + \sqrt{\Delta}}
{2 \lambda \theta (\hat{\mu} + \theta)}   \biggr) ^ {+}
\label{eq:palloc}
\end{eqnarray}
where
$
\Delta =  {\lambda}^2\,\hat{\mu}^2\,{\sigma_z}^4 +
4 \lambda \, \hat{\mu}^2 \theta {\sigma_z}^2 +
4 {\theta}^2 \hat{\mu} \lambda {\sigma_z}^2
$ and $\lambda$ determines $P_{av}$ in \eqref{eq:pconst1}.
Note that $P_d(\mhat, \theta, \lambda)  > 0$ 
for $\hat{\mu} > \lambda {\sigma_z}^2$.
This power allocation is the same as that obtained
in \cite{KleGal01ISIT}, which considers a fading
channel with constant estimation error, as opposed
to the time-varying estimation error in our model.


\begin{figure}[htbp]
\begin{center}
\includegraphics[width = 12.0cm, height = 9cm,angle=0]{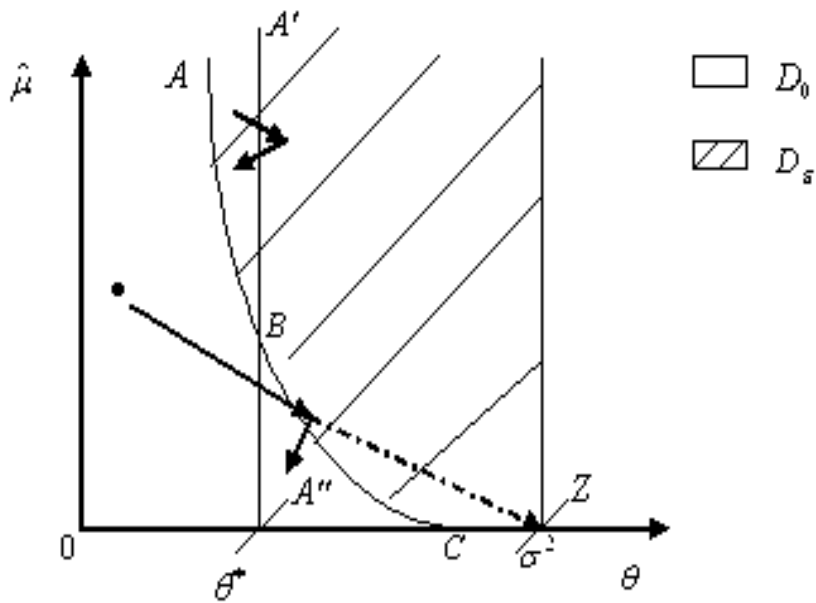}
\caption{Illustration of the dynamics of the optimal switching
policy for pilot power control.}
\label{fig:sysdyn1}
\end{center}
\end{figure}

\begin{figure}[htbp]
\begin{center}
\includegraphics[width = 12.0cm, height = 9cm,angle=0]{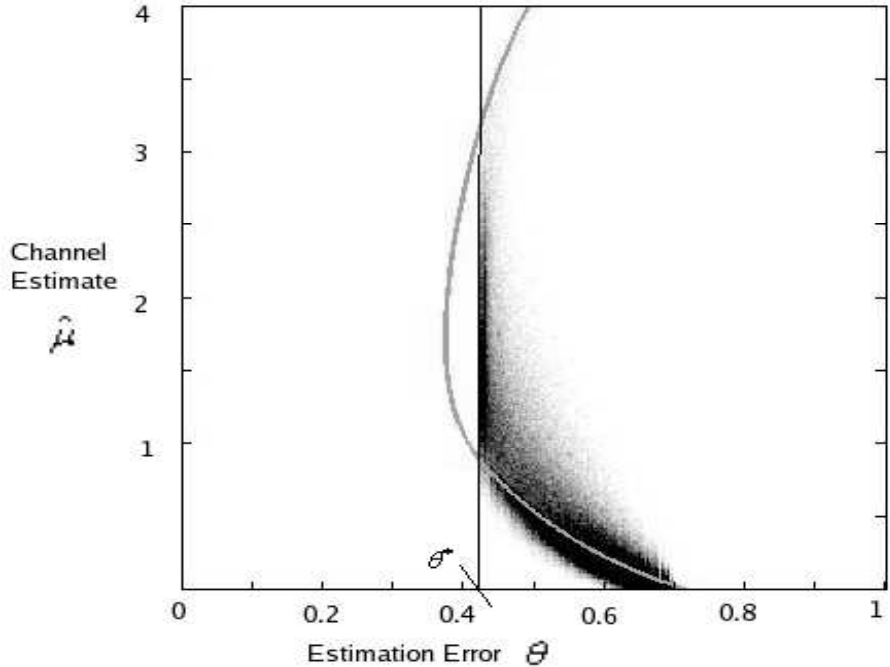}
\caption{Trace of the state $(\mhat, \theta)$ obtained by simulating 
\eqref{eq:estevol1} and \eqref{eq:errdyn} with bang-bang control.
Parameters are $M =1, N =200$, so that $dt = M/N = 0.005$,
and $\emax = 12 (10.8 dB), \rho =2, \sh =1, \sz =1$. 
A higher density of dots (darker regions) corresponds to 
higher steady-state probabilities. Also shown is the
free boundary computed via the diffusion model.
The state lies along the free boundary and 
the $\theta = \theta^\star$ line.
Also, the probability decreases with $\mhat$.}
\label{fig:trace}
\end{center}
\end{figure}

\section{Behavior of the Optimal Policy}
\label{solanal}

From Theorem \ref{thm} the optimal pilot power control
policy is determined by the switching boundary in the 
state space $(\hat{\mu},\theta)$, which is defined
by the condition $b=\lambda$.
This is referred to as a ``free boundary'' condition
for the Bellman PDE \eqref{eq:bell2} \cite{Zwilli98, Friedm82}.

The dynamical behavior of the optimal pilot power control
policy is illustrated in Fig\ \ref{fig:sysdyn1}.
The vertical and horizontal axes correspond
to the state variables $\hat{\mu}$ and $\theta$, respectively.
The shaded region, $D_{\epsilon}$, is the region of the
state space in which $\epsilon = \epsilon_{max}$,
and $\epsilon=0$ in the complementary region $D_{0}$.
These two regions are separated by the free boundary, $AC$.
The penalty factor $\lambda$
determines the position of this boundary,
and the associated value of $P_{av}$.

The vertical line $A'A''$ in the figure corresponds
to the estimation error variance $\theta^\star$,
which results from taking $\epsilon=\epsilon_{max}$
for all $t$. Clearly, in steady state the estimation
error variance cannot be lower than this value,
hence the steady-state probability density function (pdf) 
of the state $(\mhat,\theta)$
is zero for $\theta < \theta^\star$.
Substituting $\epsilon=\epsilon_{max}$
in \eqref{eq:errdyn} and setting $\frac{d\theta}{dt} = 0$ gives
\begin{equation}\label{eq:stkalerr}
\theta^\star = (\sqrt{1 + 2 \sigma_h^2 \gamma} - 1)/\gamma,
\,\, \textrm{where}\,\,\, \gamma = \epsilon_{max}/(\rho \sigma_z^2).
\end{equation}

Suppose that the initial state is in $D_0$.
With $\epsilon (t)=0$ the state evolution equations
\eqref{eq:estevol1} and \eqref{eq:errdyn} become
$d\mhat(t) = - 2\rho \mhat(t) dt$ and
$d\theta (t)/dt = 2 \rho (\sigma_h^2 - \theta (t))$.
This implies that the state trajectory is a straight line
towards the point $Z$ until it hits the free boundary,
as illustrated in Fig. \ref{fig:sysdyn1}.
Therefore, for $P_{av} > 0$, $\lambda$ must
be selected so that the point $Z$ lies in $D_{\epsilon}$.
Otherwise, the state trajectory eventually drifts to $Z$
and stays there, corresponding to 
$\epsilon=0$ for all $t$, 
$P_{av}=0$ (because $\hat{\mu}=0$), and $R=0$.
If the trajectory hits the free boundary below the point $B$,
then it is pushed back into $D_0$. This is
because at the boundary $\e(t) = \emax$
and for $\theta(t) > \theta^\star$, the drift term in 
\eqref{eq:errdyn} is negative, namely, 
$d\theta/dt = 2\rho \,({\sigma_h}^2- \theta (t)) -
\frac{\emax \theta^2 (t)}{\sigma_z^2} < 0$. 
Otherwise, if the trajectory hits the boundary above point $B$,
it continues into $D_{\epsilon}$ and settles
along the line $A'A''$ where the drift $d\theta/dt = 0$.
For the discrete-time model with small, but positive $\delta t$,
the state trajectory
zig-zags around the boundary, as shown in Fig. \ref{fig:sysdyn1}.
Hence if the free boundary $AC$ intersects $A'A''$ at
point $B$, then in steady state the probability mass
must be concentrated along the curve $A'BC$. This is
verified through Monte Carlo simulations and illustrated
in Fig. \ref{fig:trace}. Points in the state space
are shown corresponding to a realization generated from
\eqref{eq:estevol1} and \eqref{eq:errdyn} 
with $M = 1, N = 200$ 
(so that $\delta t = M/N = 0.005$).

The preceding discussion suggests that the steady-state 
probability associated with states not on the curve defined
by the free boundary and $\theta= \theta^\star$ 
tends to zero in the continuous-time limit.
This is stated formally in the next section. 
We also remark that in region $D_0$ 
the PDE \eqref{eq:bellfinal1} is a ``transport equation''
\cite{Zwilli98}, which has an analytical solution containing an
arbitrary function of a single variable. Determining
this function and the constant $C$ appears to be difficult,
so that we will take an alternative (more direct) approach to 
determining the free boundary.

\section{Steady State Behavior With Switching Policy}
\label{steadystate}

In this section we characterize the steady-state behavior 
of the state trajectory with the optimal switching 
(bang-bang) training policy, and compare with some simpler policies.
In particular, we give the first-order pdf over
the free boundary, which we subsequently use
to compute the optimized boundary explicitly.

We will denote the free boundary as $\theta_\e(\mhat)$ for $\mhat \ge 0$.
To simplify the analysis we make the following assumptions:
\begin{itemize}
\item[] (P1) \, The free boundary 
$\theta_\e(\cdot):[0, \infty) \to (\theta^\star, \sh)$ 
is a continuously differentiable curve such that
\begin{equation}\label{eq:dervcon}
\sh - \theta_\e(x) + \mhat \frac{d\theta_\e(x)}{dx} \ge 0 \qquad \textrm{for}\,\, x \ge 0.
\end{equation}
\item[] (P2)\, The function $\theta_\e (\cdot)$ is one-to-one, i.e., 
for any $x_1, x_2 \ge 0$ such that $x_1 \not= x_2$,
$\theta_\e(x_1) \not= \theta_\e(x_2)$. 
\end{itemize}
Note that (P1) requires $\emax$ to be large enough so that 
the entire free boundary ($AC$ in Fig. \ref{fig:sysdyn1}) 
lies to the right of $\theta = \theta^\star$.
(That is, they do not intersect.) The condition 
\eqref{eq:dervcon} on the derivative of the
free boundary curve is mild. Geometrically, it implies that 
the region enclosed by the free boundary $AC$ 
and point $Z = (0, \sh)$ is convex. 
This condition is indeed satisfied by the optimized free boundaries 
computed in later sections.


\begin{proposition}\label{lm:strip}
Let the pilot power as a function of the 
state $(\mhat(t), \theta(t))$ be given by
\begin{equation}\label{eq:deffbd}
\epsilon(t) = \left \{ \begin{array}{ll}
  \epsilon_{max} & \textrm{if $ \theta(t) \ge \theta_\e(\mhat(t))$}\\
  0 & \textrm{otherwise}.
\end{array} \right.
\end{equation}
Then for any $\eta >0$, the solution to 
\eqref{eq:estevol1} and \eqref{eq:errdyn} satisfies
\begin{equation}\label{eq:stripprob}
\lim_{t \to \infty} \Pr\{ |\theta(t) - \theta_\e(\mhat(t))| > \eta \} = 0.
\end{equation} 
\end{proposition}

The proof is given in Appendix \ref{ap:strip}.
The theorem implies that for large $t$ the state $(\mhat(t), \theta(t))$ 
moves along the free boundary $\{\theta_\e(\mhat)\}$.
Hence for the discrete-time system
with large $N$, the state is typically confined
to a narrow strip around the free boundary.
\footnote{This behavior of a controlled Markov process in which the 
initial state space reduces to a much smaller set under a certain class of
control inputs is called "state space collapse" \cite{Bramso98QUE}.}

\begin{theorem}\label{SSP}
Given the pilot power control \eqref{eq:deffbd}, 
the steady-state probability of
training conditioned on the channel estimate $\mhat=u$ is
\begin{equation}\label{eq:ProbTrain}
p(u) = \lim_{t \to \infty} \Pr\{\theta(t) \ge \theta_\e(\mhat)| \mhat(t) = u\}  =  \frac{2 \rho \sigma_z^2 [\sigma_h^2 -
  \theta_\e(u)]}{[\theta_\e(u)]^2\emax},
\end{equation}
and the steady-state pdf of the channel estimate $\mhat$ is
\begin{equation}\label{eq:SSgeneral}
f_{\hat{\mu}}(u) = \frac{1}{\sigma_h^2 - \theta_\e(u)}
\exp \left(- \int_0^{u}\frac{1}{\sigma_h^2 - \theta_\e(s)}ds
\right), \qquad u \ge 0 .
\end{equation}
\end{theorem}

The proof is given in Appendix \ref{ap:ssp}.
From \eqref{eq:ProbTrain}
the average training power for the 
pilot power control scheme can be computed as
\begin{equation}\label{eq:TrainGeneral}
\epsilon_{avg} = \int_0^\infty \frac{2 \rho \sigma_z^2 (\sigma_h^2 -
  \theta_\e(u))}{[\theta_\e(u)]^2} f_{\hat{\mu}}(u) \, du.
\end{equation}
Therefore if $\epsilon_{max}$ is large enough so that
$\theta^\star < \theta_\e(u)$ for all $u \ge 0$, then
neither the pdf $f_{\hat{\mu}}(\cdot)$ 
nor the average training power $\epsilon_{avg}$
depends on $\epsilon_{max}$.
This is because as $\emax$ increases,
the probability of training, given by \eqref{eq:ProbTrain}, 
decreases so that the average training power given $\mhat$,
namely $\emax p(\mhat)$, remains unchanged\footnote{This
ignores the overhead due to the insertion of training symbols.
According to the subsequent discussion in Sec. \ref{tl}, 
this overhead is reduced by increasing $\emax$.
However, for the diffusion approximation
to be accurate, $\emax \frac{M}{N}$ must be small,
hence $\emax$ cannot be too large.}.
In addition, we observe that $f_{\hat{\mu}} (\cdot)$ 
is independent of the correlation parameter $\rho$. 

Now consider the case in which the
free boundary is constrained to be vertical, that is,
$\theta(\mhat) = \theta_v, \,\, \forall \mhat \ge 0$.
This still corresponds to a switching policy,
but where the variance of the
channel estimation error is constrained to be a constant,
independent of the channel estimate.
From \eqref{eq:SSgeneral} the steady-state
pdf of $\mhat$ is exponential, i.e.,
$f_{\mhat}(u) = \frac{1}{\sh - \theta_v}
\exp(-\frac{u}{\sh - \theta_v})$,
and the average training power is
$\epsilon_v = \frac{2 \rho \sigma_z^2 
(\sigma_h^2 - \theta_v)}{\theta_v^2}$. 

Now consider the \emph{constant} pilot power control policy,
where $\epsilon(t)= \epsilon_v$ (constant) for all $t$. 
Substituting $\epsilon = \epsilon_v$
into \eqref{eq:errdyn} and setting $\frac{d\theta}{dt} = 0$ 
implies that the steady-state estimation error variance 
$\theta(t)=\theta_v$ for all $t$.
In addition, the steady-state pdf of $\mhat$
is exponential with mean $\sh - \theta_v$. 
Hence for a given average training power,
constant pilot power can give {\em exactly} 
the same estimation error and steady-state pdf as
the switching policy with a vertical boundary.
Both schemes therefore achieve the same rate with
a total power constraint (ignoring overhead
due to pilot insertion). We will see that the
optimized boundary is \emph{not} vertical, which
implies that adaptive pilot power control can
perform better than constant pilot power.


We also observe that the same performance as the optimal
switching policy can be achieved
by {\em continuously} varying the training power 
as a function of $\mhat$. Namely, taking
$\epsilon(\mhat) =  \frac{2 \rho \sigma_z^2 
(\sigma_h^2 - \theta(\mhat))}{\theta(\mhat)^2}$
gives the same steady-state pdf
and training power as in \eqref{eq:SSgeneral} and 
\eqref{eq:TrainGeneral}, respectively. 
However, this scheme corresponds to feeding back
the pilot power as a sequence of real numbers, 
which in principle require infinite precision.
In contrast, the switching policy can be implemented by
fixing the training power and
varying the rate at which pilot symbols are inserted.
The transmitter therefore does not need to know 
the exact value of the channel estimate.

More specifically, the optimal switching policy inserts 
pilots of power $\epsilon_{max}$ with probability
$q = \frac{2 \rho \sigma_z^2 (\sigma_h^2 -
\theta(u))}{\epsilon_{max}\theta(u)^2}$ 
(or equivalently,
once every $1/q$ coherence blocks) 
when the channel estimate $\mhat=u$.
This requires at most one bit per coherence block to inform
the transmitter whether or not to train in 
the next block. (Of course, the feedback
can be substantially reduced by exploiting channel correlations.)
The switching policy therefore requires fewer
training symbols than continuous pilot power control,
which requires a pilot symbol every coherence block.

\section{Free Boundary with Optimal Data Power Allocation}
\label{water}
From the preceding discussion the optimal pilot policy
is determined by the free boundary. Here we compute the free boundary
by observing that this boundary must maximize the rate objective,
assuming a switching policy for the pilot power.
A difficulty is that the rate objective of interest
is $N R(P/N,\mhat,\theta)$, whereas the steady-state probabilities 
in Theorem \ref{SSP} were derived in the limit as $N \to \infty$. 
In what follows we use the asymptotic probabilities 
in Theorem \ref{SSP} to approximate the steady-state probabilities 
corresponding to large but finite $N$. 
Simulation results have shown that the resulting free boundary
is insensitive to the choice of $N$ in the objective.
Also, subsequent simulation results in Section \ref{wf:num} show 
that the analytical performance results accurately predict the 
performance of the corresponding discrete-time model 
with the optimal switching policy when $N$ 
is a few hundred.
\subsection{Analytical Solution}
With the preceding approximation for large but finite $N$
the optimal free boundary can be computed as the solution to the 
following functional optimization problem,  
\begin{eqnarray}
\label{eq:fbopt}
\begin{aligned}
\max_{P(u), \theta_\e(u)}\,  &
 \,\,\int_{0}^{\infty} N R[P(u)/N, u, \theta_\e(u)] 
f_{\mhat}(u) du   \\
\text{subject to:} & \label{eq:pconst_ss} \int_{0}^{\infty} 
P(u) f_{\mhat}(u)du +
\int_{0}^{\infty} \epsilon(u) \,f_{\mhat}(u)\,du \le
\,P_{av},\\
\text{and} & \qquad \qquad \theta_\e(u) \ge \theta^\star \quad \text{for}
\,\, u \ge 0,
\end{aligned}
\end{eqnarray}
where $\epsilon(u) = \frac{2 \rho \sigma_z^2 
[\sh - \theta_\e(u)]}{[\theta_\e(u)]^2}$ 
can be interpreted as the average training power 
when the estimate $\mhat =u$. 
The objective is the achievable rate averaged over
the free boundary since for large $N$, the entire probability mass
becomes concentrated \emph{on} the boundary.

To proceed, define the Lagrangian function as 
\begin{equation}\label{eq:LagFunc}
L_{\lambda} [P, u, \theta] = 
N R(P/N, u, \theta) - \lambda \left(P + \frac{2 \rho \sigma_z^2 
(\sh - \theta)}{\theta^2} \right)
\end{equation}
Analogous to \eqref{eq:lag}, the optimization problem 
\eqref{eq:fbopt} can be re-stated as
\begin{equation}\label{eq:fbd_lag}
\max_{P(u), \theta_\e(u)}\,  
 \,\,\int_{0}^{\infty} L_{\lambda}[P(u), u, \theta_\e(u)] f_{\mhat}(u)du  
\end{equation}
such that $\theta_\e(u) \ge \theta^\star$ and $\lambda$ is chosen
to satisfy the power constraint \eqref{eq:fbopt} with equality.
It is shown in Appendix \ref{prop-wf-fbd} that the solution is given by
\begin{equation}
\label{eq:fb1}
\theta_\e^\star(u) = \max \{\theta^\star, \theta_f(u)\},
\end{equation}
where $\theta^\star$ is defined in \eqref{eq:stkalerr},
and $\theta_f(u)$ satisfies
\begin{multline}\label{eq:analytic_fbd}
(\sh - \theta_f(u))\frac{\partial L_\lambda}{\partial \theta}
[N P_d(u, \theta_f(u), \lambda), u, \theta_f(u)] +
L_\lambda[NP_d(u, \theta_f(u), \lambda),
u, \theta_f(u)] = \\
\int_{u}^\infty L_\lambda[NP_d(v, \theta_f(v), \lambda),
v, \theta_f(v)] \frac{\exp \left(- \int_{u}^v\frac{1}{\sh -
      \theta_f(s)}ds\right)}{\sh - \theta_f(v)}dv
\end{multline}
for $ \,\, u \ge 0$, where
\begin{equation}\label{eq:PratialLag}
\frac{\partial L_{\lambda}}{\partial \theta} [P, u, \theta] =
-\frac{N\,P^2 u}{(P\theta + P u + N\sigma_z^2)(P\theta+ N\sigma_z^2)}
+ \frac{2 \lambda \rho \sigma_z^2(2\sigma_h^2 - \theta)}{\theta^3}
\end{equation}
and $P_d(.,.,.)$ is given by \eqref{eq:palloc}.
The optimal data power allocation is given by 
$P^\star(u) = N P_d(u, \theta_\e^\star(u), \lambda)$.

The condition \eqref{eq:analytic_fbd} gives the value
$\theta_f(u)$ as a functional of the free boundary
$\theta_f(x)$ for $x > u$. Hence we can compute the 
boundary numerically via a backward recursion
provided that $\theta_f(u)$ is known for large values of $u$.
Moreover, it can be shown that $\theta_f(u)$ is a decreasing 
function of $u$ and as $u \to \infty$, 
it converges to a constant value, that is,  
$\theta_f(u) \to \theta_{\infty}$. 
Taking the limit $u \to \infty$ on both sides of 
\eqref{eq:analytic_fbd}, this value can be shown to satisfy
\begin{equation}
\frac{2\lambda \rho \sz (2\sh -\theta_{\infty})}{\theta_{\infty}^2} = 
N \frac{\sqrt{1+ \frac{4\theta_{\infty}}{\lambda\sz}} -1}{\sqrt{1+ 
\frac{4\theta_{\infty}}{\lambda\sz}} +1} 
\end{equation}
Hence as long as $\epsilon_{max}$ is large enough so that 
$\theta^\star < \theta_{\infty}$, the free boundary
is given by $\theta_f(u)$, independent of $\epsilon_{max}$. 

Substituting $u = 0$ into \eqref{eq:analytic_fbd}, 
and using the fact that $\theta_\e^\star(u) = \theta_f(u)$ 
for all $u \ge 0$ 
(since $\theta^\star < \theta_{\infty}$), and $P^\star(0) = 0$,
we obtain
\begin{equation}\label{eq:theta_not}
\frac{[\sh - \theta_\e^\star(0)]^2}{[\theta_\e^\star(0)]^3} = \frac{1}{4
  \rho \sigma_z^2} \left[\frac{\bar{R}}{\lambda} - P_{av}\right].
\end{equation}
where $\bar{R}$ is the optimized objective in \eqref{eq:fbopt}.
Note that this relation depends on $N$ only through 
the water-filling level $\lambda$. Clearly, 
the left-hand side of \eqref{eq:theta_not} is positive, 
which implies that we should have
$\bar{R} \ge \lambda P_{av}$. Also, as $P_{av} \to 0$
(low SNRs), $\bar{R} \to 0$ and $\lambda$ increases.
Similarily, the right-hand side decreases to zero
as $\rho \to \infty$ (fast fading), so that in
both cases $\theta^\star(0) \to \sh$.
  


\subsection{Numerical Approach to Free Boundary Problem}
The preceding approach to computing the optimal free boundary
relies on the asymptotic pdf of the state in Theorem \ref{SSP}.
Alternatively, it is possible to solve the continuous-time
Bellman equation \eqref{eq:bellfinal1} directly.
This is potentially useful for other scenarios in which 
the steady-state distribution is more difficult to obtain.
A challenge, however, is that the optimized free boundary 
is unknown {\em a priori}, i.e., it is obtained as part
of the solution. Hence none of the standard numerical
methods for solving PDEs, which rely on specified boundary conditions,
can be directly applied.

It is shown in Appendix \ref{VarInq} that a numerical solution
to the free boundary Bellman equation can be obtained
by re-formulating the problem as a quadratic program.
That method can be used to obtain a solution to a 
general class of free boundary problems and 
does not require knowledge of the steady-state statistics. 
However, such a numerical computation has high complexity,
and can be sensitive to parameter variations.
In particular, the free boundary obtained from that
method is often irregular (not smooth)
due to discretization and finite-precision effects.

\subsection{Numerical Results}
\label{wf:num}

Here we present some numerical examples of free boundaries
obtained by solving the optimization problem \eqref{eq:fbopt}
along with performance results. The analytical (diffusion) results
are also compared with results from a Monte Carlo simulation
of the discrete-time system. To solve \eqref{eq:fbopt}
we discretize the $\mhat$ axis and also truncate it 
at a value $U_T >> \sh$.
For all of the results in this section
$\sh =\sz =1$, and $\epsilon_{max} = 15$ (11.76 dB).
The SNR is then $\frac{P_{av}\sh}{\sz} = P_{av}$.

\begin{figure}[htbp]
\begin{center}
\includegraphics[width = 12.0cm, height = 9cm, angle=0]{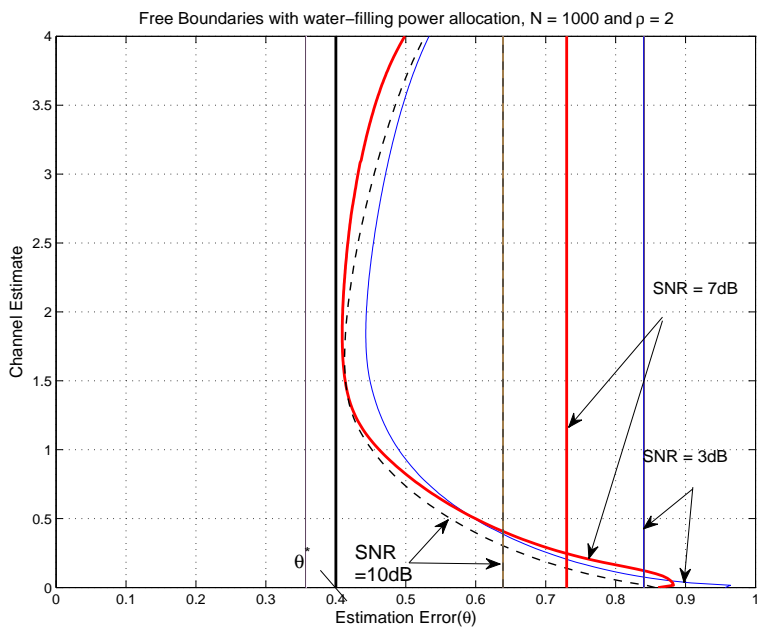}
\caption{Optimal free and vertical boundaries 
with the water-filling data power allocation ($N=1000$, $\rho=2$).}
\label{fig:fbdN1000}
\end{center}
\end{figure}

\paragraph{Free Boundary Examples} 
Fig. \ref{fig:fbdN1000} shows free boundaries 
corresponding to $N = 1000$ and $\rho =2$ 
for different SNRs. The channel correlation with lag 1000
is therefore $e^{-2}= 0.135$, corresponding to
relatively fast fading. (We abuse notation in this
section by referring to $\mhat$ as a particular realization 
of the channel estimate.) Also shown are the
optimized vertical boundaries with a switching policy
(i.e., $\theta(u) = \theta_v, \,\,\, \forall u \ge 0$
with optimized $\theta_v$).
Recall that the performance with the vertical boundary
is the same as training with a constant fraction of power,
which results in the estimation error $\theta_v$.
For each boundary the data power allocation is given
by the optimal water-filling power allocation
in \eqref{eq:palloc}.

The free boundaries are shaped so that the estimation error
is larger for small values of $\mhat$, and smaller for
larger values of $\mhat$. The reason for this 
is that the pdf of $\mhat$ is larger when $\mhat$ is 
close to zero where the instantaneous rate $R(\cdot)$ is small. 
(In fact, $R=0$ for $\mhat \le \lambda \sz$.)
Allowing larger estimation errors for small $\mhat$ 
(relative to the vertical boundary) therefore does 
not significantly reduce the overall ergodic rate, 
whereas it saves a significant amount of training power. 
Furthermore, shifting the savings in training power to 
larger values of $\mhat$ reduces the estimation error 
for those values, thereby increasing the rate
(since the rate increases with $\mhat$). 
It will be shown in Sec \ref{onoff:analytic} that 
with an optimized on-off power allocation,
for large enough $N$ the achievable rate for the free boundary
control depends on the shape of the free boundary only through
the harmonic mean of the function $\sh - \theta_{\e}(x)$. 
This can also be used to show that the boundary 
has the general shape shown in Fig. \ref{fig:fbdN1000}.

Of course, the estimation error cannot be too large
for small $\mhat$, since otherwise $\mhat$ may decreases to zero.
(Note from \eqref{eq:SSgeneral} that as 
$\theta_{\e}(\mhat) \to \sh$, the
density $f_{\mhat} (u)$ becomes concentrated around $\mhat = 0$.)
Also, Fig. \ref{fig:fbdN1000} shows that
at smaller SNRs $\theta_{\e}(0)$ is closer to $\sh$,
so that the free boundary is more skewed.
The curvature of the boundary near $\mhat = 4$ is 
a numerical artifact due to truncation of the boundary at $U_T=5$.
Namely, this curvature disappears as $U_T$ increases,
since as discussed in the last section, $\theta_f (\mhat)$
is a decreasing function of $\mhat$ and approaches
the value $\theta_{\infty}$ as $\mhat$
becomes large.

\begin{figure}[htbp]
\begin{center}
\includegraphics[width = 12.0cm, height = 9cm, angle=0]{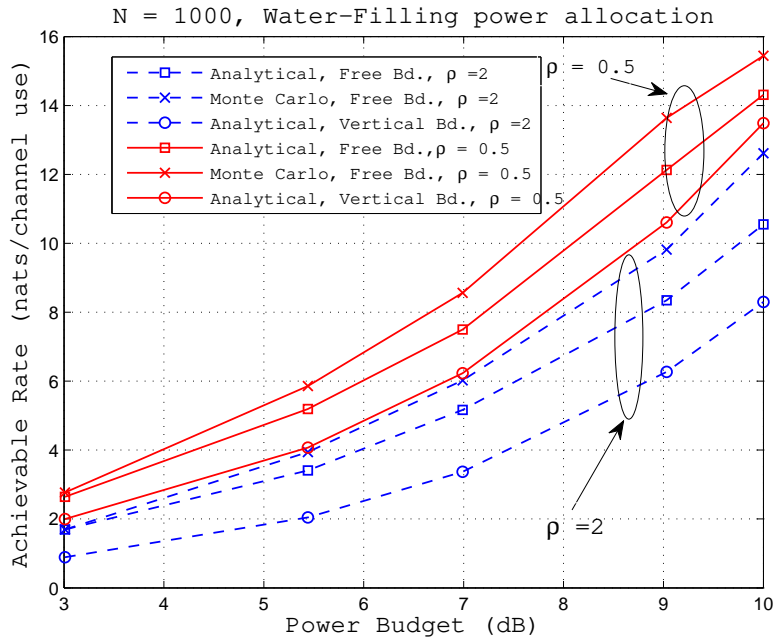}
\caption{Achievable rate versus SNR with optimal free and vertical 
boundary pilot power control.}
\label{fig:N1000_Cap}
\end{center}
\end{figure}

\paragraph{Gain in Achievable Rate}
Fig. \ref{fig:N1000_Cap} compares the rate objective
with the optimized free boundary to that achieved 
with the optimized vertical boundary at different SNRs. 
With $\rho = 2$ (fast fading) these results show that 
optimized pilot power control gives substantial gains 
at low SNRs (e.g., a factor of two at an SNR of $3~dB$).
The percentage gain diminishes with the SNR.
With $\rho=0.5$, corresponding to a correlation
of 0.6 with lag 1000, the gain in achievable rate is relatively small.
The optimized free boundary gives the most gain at low SNRs
and fast fading, since in that region the training power 
constitutes a larger percentage of the total power budget.

\paragraph{Comparison with Monte Carlo Simulations}
Fig. \ref{fig:N1000_Cap} also compares the analytical
results from the diffusion model with the performance
of the original discrete-time model obtained from Monte Carlo simulations.
Specifically, the discrete-time system \eqref{eq:model}-\eqref{eq:gm}
was simulated with the
Kalman filter estimator \eqref{eq:kalest}-\eqref{eq:condvar}
and the switching policy for pilot power defined by the optimized 
free and vertical boundaries.\footnote{The simulated results
assume the optimized boundaries obtained in the diffusion limit,
since the optimized boundary for the discrete-time system, given
by the solution to \eqref{eq:bell1}, is much more difficult to compute.
Additional simulation results have shown that the solution
to \eqref{eq:bell1} is quite close to the asymptotic (diffusion)
boundary, and gives essentially the same performance.}
The comparison in Fig. \ref{fig:N1000_Cap} shows that the 
analytical results for the optimized free boundary 
underestimates the achievable rate
by about 15-20\%. The simulation and numerical optimization 
give nearly the same values with the optimized vertical boundary.

In addition to the parameters $N$ and $\rho$, which are the
same as for the analytical results,
for the discrete-time model another parameter is $M$, the number
of samples per coherence block. Recall that the diffusion model
is obtained in the limit as $M/N \to 0$, hence for fixed $N$
the analytical results should be more accurate for smaller values of $M$
(corresponding to higher correlations between successive channel gains).
However, smaller values of $M$ incur more overhead, 
since the training and channel state feedback occur 
each coherence block. (We discuss this further in Sec. \ref{tl}.)
The results in Fig. \ref{fig:N1000_Cap} correspond to $M=5$.

\begin{figure}[htbp]
\begin{center}
\includegraphics[width = 12.0cm, height = 9cm, angle=0]{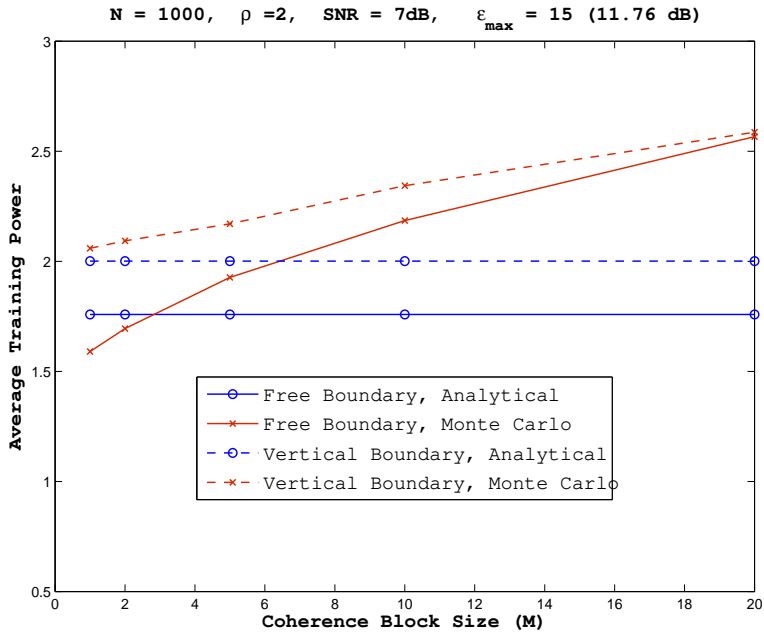}
\caption{Comparison of training power values obtained from Monte Carlo 
simulations and analysis with different values of $M$.
($N =1000$, $\rho =2$ and $SNR = 7dB$)}
\label{fig:N1000_Eavg_diffM}
\end{center}
\end{figure}

\begin{figure}[htbp]
\begin{center}
\includegraphics[width = 12.0cm, height = 9cm, angle=0]{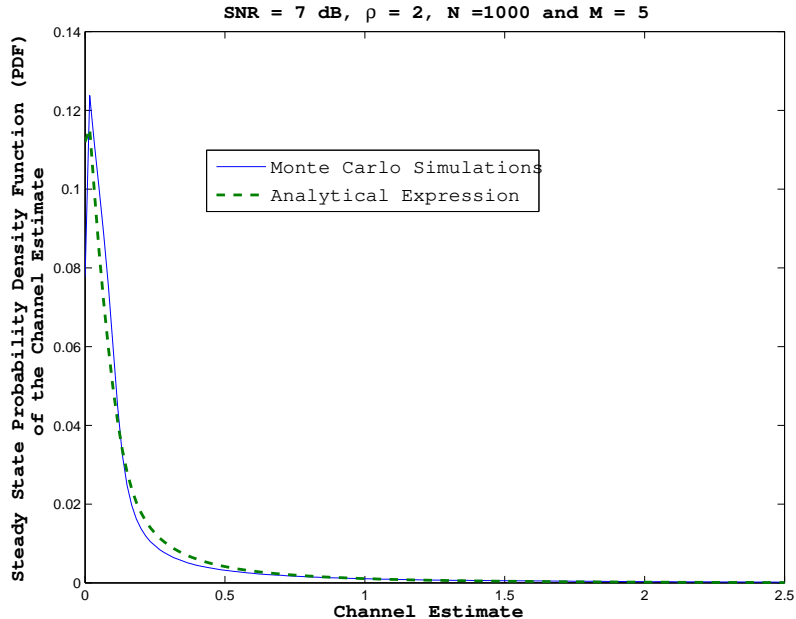}
\caption{Channel estimate pdf $f_{\mhat}(u)$ obtained 
from Monte Carlo simulations and the diffusion analysis.}
\label{fig:N1000_sspdf}
\end{center}
\end{figure}

To see the effect of varying $M$ with fixed $N$,
Fig. \ref{fig:N1000_Eavg_diffM} compares the
optimized average training power obtained via analysis
and Monte Carlo simulations for different values of $M$
with an SNR of 7 dB. As expected, the two curves
grow apart as $M$ increases, but are reasonably
close for $M \le 5$.
Fig \ref{fig:N1000_sspdf} compares the simulated steady-state pdf of the 
channel estimate $\mhat$ with $M = 5$ with the
asymptotic pdf \eqref{eq:SSgeneral}.
The two curves nearly overlap.
Further results show that achievable rates computed from the
analytical model nearly match the simulated rates
with the vertical boundary over a wide range of $M$, 
whereas for the optimized boundary
the difference remains similar to that shown in 
Fig. \ref{fig:N1000_Cap}.

\begin{figure}
    \centering
    \subfigure[Training Power]
    {\label{fig:power}
    \includegraphics[width = 12.0cm, height = 9cm, angle=0]{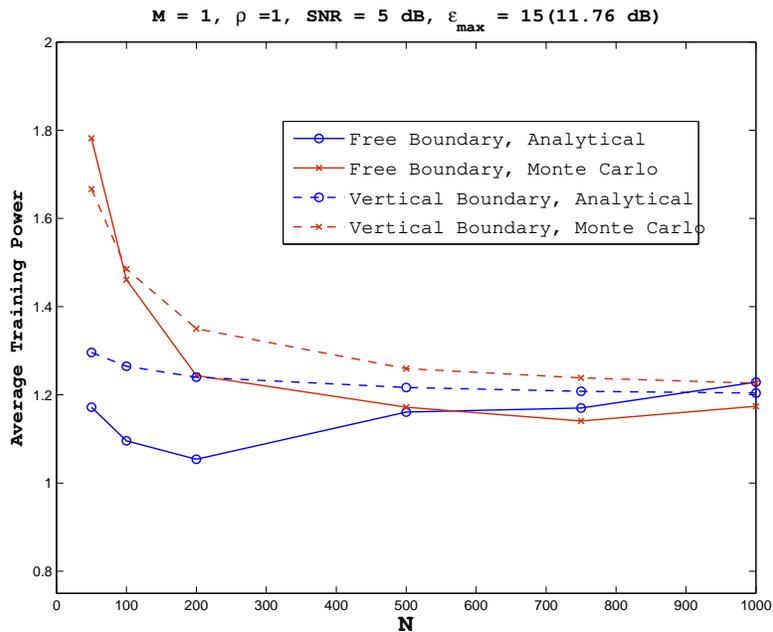}}
    \subfigure[Achievable Rate]
    {\label{fig:rate}
    \includegraphics[width = 12.0cm, height = 9cm, angle=0]{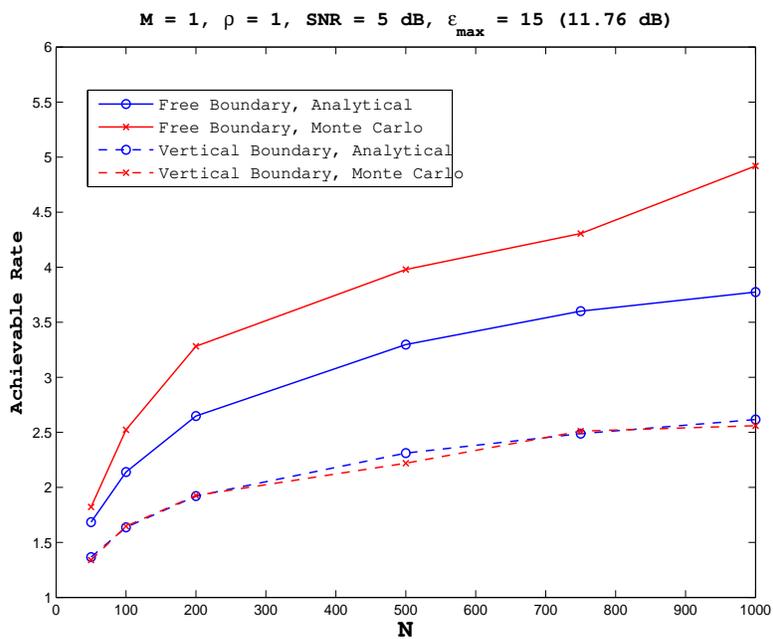} }
    \caption{Comparison of training power and achievable rates 
obtained from Monte Carlo simulations 
and the diffusion analysis with different values of $N$. 
($M =1$, $\rho =1$, $SNR = 5~dB$)}
\label{fig:Cap_diffN}
\end{figure}

Fig. \ref{fig:Cap_diffN} compares analytical and simulated
results as a function of $N$ with $M=1$.
For a fixed $P_{av}$ and $\rho$, as $N$ decreases,
the SNR per sub-channel increases and the channel varies
at a faster rate since the correlation across
$N$ channel uses is fixed at $e^{-\rho}$. 
The achievable rate increases with $N$
due to the increase in training and channel state feedback.
(In Sec. \ref{on-off} we show that the rate increases as $\log N$.)
Again the analytical results
closely match the simulated results with the optimized
vertical boundary, and underestimate the 
achievable rate by $15-20\%$ with the 
optimized free boundary.\footnote{The gap between
the analytical and simulation results with the optimized
free boundary is due to the fact that as $N\to\infty$,
the scaled rate objective increases without bound.
This occurs even though from Theorem \ref{SSP}
the steady-state distribution of the system state 
converges to the exponential distribution.}
The plots for average training power
become close for $ N \ge 200$.


%




\section{On-Off Data Power Allocation}
\label{on-off}
An important consequence of the optimal switching policy
for pilot power control is that it requires no more than
one bit feedback per coherence block.
However, optimal {\em data} power control
still requires infinite-precision feedback.
Therefore to reduce the overall feedback rate, we now 
consider on-off data power allocation, which also requires
at most one bit feedback per coherence block.
(The feedback could be reduced further by exploiting 
the time correlation of the channel.)
For the optimization problem \eqref{eq:fbopt} we therefore set
\begin{equation}\label{eq:ponoff}
P(u) = \left \{ \begin{array}{ll}
  P_0 & \textrm{if $ u > \mhat_{0}$}\\
  0 & \textrm{otherwise},
\end{array} \right.
\end{equation}
where the threshold $\mhat_0$ will be optimized,
and restate \eqref{eq:fbopt} as

\begin{eqnarray}
\label{eq:fbopt-onoff}
\begin{aligned}
\max_{P_0, \mhat_0, \theta_{\e}(u)}\,  &
 \,\,\int_{\mhat_0}^{\infty} N R(P_0/N, u, \theta_{\e}(u)) f_{\mhat}(u)du   \\
\text{subject to:} & \qquad\label{eq:pconst2_ss}  q P_0   +
\epsilon_{avg}\le \,P_{av},\\
\text{and} & \qquad  \theta_{\e}(u) \ge \theta^\star, \quad
\text{for} \,\,\, u \ge 0,
\end{aligned}
\end{eqnarray}
where $f_{\mhat}(u)$ is the steady-state pdf of the
channel estimate given by \eqref{eq:SSgeneral}, 
$\epsilon_{avg} = \int_{0}^{\infty} \epsilon(u) \,f_{\mhat}(u)\,du$
is the average training power and 
$q = \Pr \{\mhat > \mhat_0\} = \int_{\mhat_0}^{\infty}  \,f_{\mhat}(u)\,du$.

\subsection{Harmonic Mean Objective}
\label{onoff:analytic}

Given a free boundary $\theta_{\e}(u)$ for $u \ge 0$, 
where $\theta_{\e}(u) \in (\theta^\star,\sh)$,
the objective in \eqref{eq:fbopt-onoff} can be re-written as
\begin{equation}\label{eq:onoff-obj}  
R_{\textrm{on-off}} (\mhat_0) =  \int_{\mhat_0}^{\infty} 
N \log\left(1+ \frac{(P_{av} - \epsilon_{avg})u}
{(P_{av} - \epsilon_{avg})\theta_{\e}(u) + \sz N q}\right) 
f_{\mhat}(u) du,
\end{equation}
where we have used the fact that the total power constraint is 
satisfied with equality. Using \eqref{eq:SSgeneral}, 
the probability of data transmission is
\begin{equation}
q = \int_{\mhat_0}^{\infty}  \,f_{\mhat}(u)\,du
= \exp\left(-\int_0^{\mhat_0} \frac{1}{\sh - \theta_\e(u)} du\right).
\end{equation}
Furthermore, the rate in \eqref{eq:onoff-obj} can be bounded as
\begin{equation}\label{eq:onoff-bounds}
  Nq \log\left(1+ \frac{(P_{av} - \epsilon_{avg})\mhat_0}
{(P_{av} - \epsilon_{avg})\sh + \sz N q}\right) \le
R_{\textrm{on-off}} (\mhat_0) \le
  Nq \log\left(1+ \frac{(P_{av} - \epsilon_{avg})(\mhat_0+\sh)}
{ \sz N q}\right),
\end{equation} 
where the upper bound follows by replacing $\theta_\e(\mhat)$ by $0$,
using Jensen's Inequality \cite{Rudin86}, and the fact that 
$\int_{\mhat_0}^{\infty}  u \,f_{\mhat}(u)\,du \le \mhat_0+\sh$.

Observing that $q$ is a decreasing function of $\mhat_0$, 
it can be shown that the upper bound  \eqref{eq:onoff-bounds} 
is maximized with threshold $\mhat_0^\star$ such that
\begin{equation}\label{eq:opt-thres1}
\int_{0}^{\mhat_0^\star} \frac{1}{\sh - \theta_\e(u)} du = 
\log\left(\frac{N}{(\log N)^{1+\delta}}\right)
\end{equation}
where $\delta \in (0,1)$ is an increasing function of $N$. 
(An exact description of $\delta$ is unecessary for
the following analysis.)
We can re-write \eqref{eq:opt-thres1} as
\begin{equation}\label{eq:opt-thres2}
\mhat_0^\star = H(\mhat_0^\star) 
\log\left(\frac{N}{(\log N)^{1+\delta}}\right)
\end{equation}
where $H(\mhat_0^\star) = 
\frac{\mhat_0^\star}{\int_{0}^{\mhat_0^\star} \frac{1}{\sh - \theta_\e(u)} du}$
is the {\em harmonic mean} 
of $\sh - \theta_\e(u)$ for $u \in [0, \mhat_0^\star]$,
that is, along the free boundary
truncated at the threshold value $\mhat_0^\star$. 
Since $H(\mhat_0^\star) \in (0, \sh)$, \eqref{eq:opt-thres2} 
implies that $\mhat_0^\star$ grows as $\log N$. 
Substituting $\mhat_0^\star$ into 
\eqref{eq:onoff-bounds}, we observe that the
upper and lower bounds have the same asymptotic growth rate,
so that the rate \eqref{eq:onoff-obj} also has this growth rate,
given by
\footnote{The notation $F_1(N) \asymp F_2(N)$ implies
that $\lim_{N \to \infty} \frac{F_1(N)}{F_2(N)} = 1$.}
\begin{eqnarray}
R_{\textrm{on-off}}(\mhat_0^\star) &\asymp& (\log N)^{1+\delta} 
\log \left(1 + \frac{(P_{av} - \epsilon_{avg}) \mhat_0^\star}{\sz (\log N)^{1+\delta}}\right)\\
&\asymp& \frac{(P_{av} - \epsilon_{avg})\mhat_0^\star}{\sz}\\
&\asymp& \frac{(P_{av} - \epsilon_{avg})H(\mhat_0^\star)}{\sz}  \log N
\end{eqnarray}  
Since $\mhat_0^\star$ maximizes the upper bound in 
\eqref{eq:onoff-bounds}, this is the growth rate 
of the achievable rate. 

We observe that this $\log N$ growth in achievable rate
is the same as the growth in achievable rate for
parallel Rayleigh fading channels (in frequency or time) 
with a sum power constraint and perfect
channel knowledge at the transmitter (e.g., see \cite{SunHon08IT}).
This is because the coherence blocks correspond to
separate degrees of freedom (i.e., the transmitter
can choose whether or not to transmit over each block),
and the number of coherence blocks increases linearly with $N$.
For our model the associated constant is
$(P_{av} - \epsilon_{avg})H(\mhat_0^\star)$, which
accounts for channel estimation error, and
depends on the channel correlation $\rho$.
This product therefore determines the shape of the free boundary. 
(Note also that $\mhat_0^\star$ depends on the free boundary.)
Namely, choosing boundary points closer to 
$\sh$ reduces $\epsilon_{avg}$, but also reduces the
harmonic mean $H(\mhat_0^\star)$, and vice versa.
The optimal boundary balances $\epsilon_{avg}$
and $H(\mhat_0^\star)$ by shifting training power
from small values of $\mhat$ to larger values, as
discussed previously in Sec. \ref{wf:num}.


\subsection{Numerical Example}
\begin{figure}[htbp]
\begin{center}
\includegraphics[width = 12.0cm, height = 9cm, angle=0]{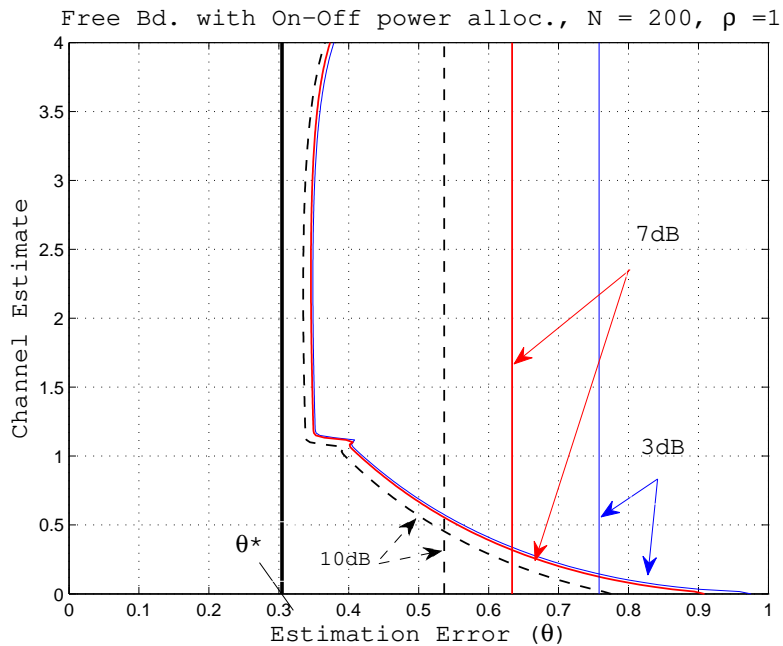}
\caption{Optimal free and vertical boundaries with 
on-off data power allocation.}
\label{fig:fbdonoff}
\end{center}
\end{figure}

Fig. \ref{fig:fbdonoff} shows free boundaries at different SNRs 
obtained by solving the optimization problem \eqref{eq:fbopt-onoff}
numerically for $N = 200$ and $\rho =1$.
Also shown are the optimized vertical boundaries
with on-off data power control. 
As with water-filling, the free boundary 
is shaped to save training power when $\mhat$ is small 
(high probability region) and re-distribute it to the 
instances when $\mhat$ is large (low probability region). 
The boundaries shown here are more irregular, 
due to the discontinuous data power allocation.
The shape of the boundary for $\mhat>\mhat_0^\star$ 
is a straight line, but does not affect the objective
since the rate depends on the harmonic mean for $\mhat \le \mhat_0^\star$. 
 


\begin{figure}[htbp]
\begin{center}
\includegraphics[width = 12.0cm, height = 9cm, angle=0]{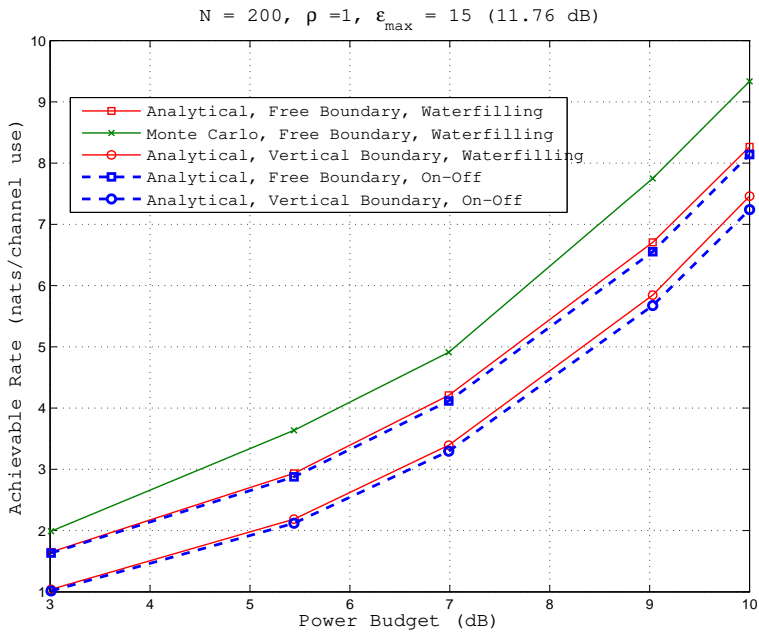}
\caption{Achievable rate versus SNR with on-off data power allocation
and optimized free and vertical boundaries for pilot power control.
Also shown for comparison are the corresponding results with 
the water-filling data power allocation.}
\label{fig:N200_Cap}
\end{center}
\end{figure}

Fig. \ref{fig:N200_Cap} shows plots of achievable rates 
versus SNR with the optimized free and vertical boundaries 
and on-off data power control. Plots corresponding to
the optimal waterfilling data power allocation are also shown for comparison. 
These results show that the performance with
the optimized on-off power allocation are nearly the same 
as with water-filling.
Also shown are the rates obtained via Monte Carlo simulations
of the discrete-time system with the optimized boundary.
Those are again higher than the rates calculated
from the diffusion model, whereas the simulated
rates with the vertical boundary closely match the analytical results.


\section{Training Symbol Overhead} 
\label{tl}

So far we have ignored the time overhead due to the 
channel uses that are occupied by the training symbols. 
Here we restate the pilot power control problem 
taking this overhead into account. 
A switching policy for the pilot power
requires that one of the $M$ channel uses in
a coherence block is a training symbol 
whenever the transmitter is directed to train.
If the channel estimate for the coherence 
block is $\mhat$, then the probability of training 
(as discussed in Sec. \ref{steadystate})
is given by $\frac{\epsilon(\mhat)}{\emax}$, where
$\epsilon(\mhat) = \frac{2\rho\sz(\sh - \theta_\e(\mhat))}{\theta_\e(\mhat)}$.
Therefore the original optimization problem \eqref{eq:fbopt} 
can be reformulated, taking the training overhead in to account,
by replacing the rate objective with
\begin{equation}
 \max_{(P(u), \theta(u))}\, 
 \,\, \int_{0}^{\infty} \left(1-\frac{\epsilon(u)}{\emax M}\right)
N R[P(u)/N, u, \theta(u)] f_{\mhat}(u)du  .
\label{eq:thpt}
\end{equation}
Of course, if either $\epsilon_{max}$ or $M$ is large, 
then the training symbol overhead is negligible and the problem
reduces to \eqref{eq:fbopt}. Otherwise, the overhead term will
influence the free boundary and ergodic rate. Specifically,
it will reduce the optimal training power $\epsilon(\mhat)$
(so that the boundary shifts towards $\theta= \sh$),
since the overhead penalty is proportional to the training power. 

\begin{figure}[htbp]
\begin{center}
\includegraphics[width = 12.0cm, height = 9cm, angle=0]{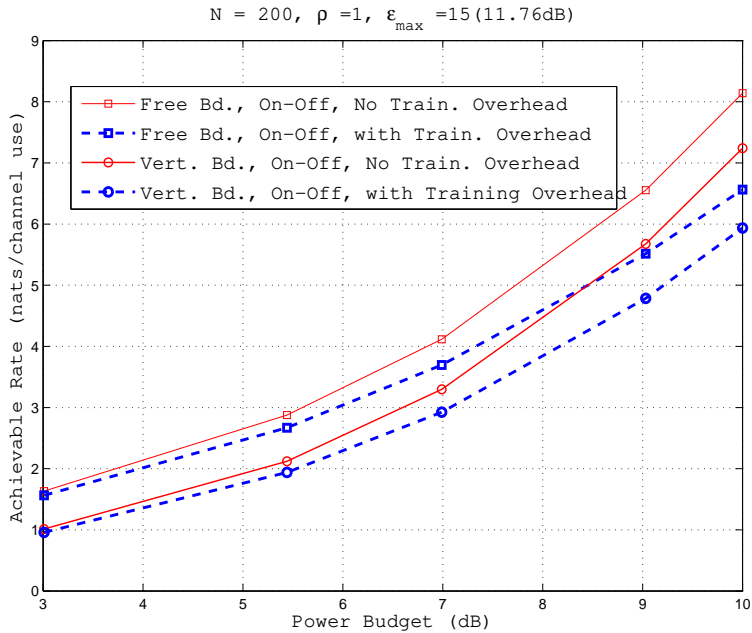}
\caption{Achievable rates taking pilot symbol overhead into account.}
\label{fig:N200_Cap_TL}
\end{center}
\end{figure}

Fig. \ref{fig:N200_Cap_TL} shows plots of 
the rate objective in \eqref{eq:thpt}
versus SNR with optimized free and vertical boundaries.
For this figure $N = 200$ and $M =1$, corresponding
to a worst-case loss in throughput due to training overhead.
Also, $\rho =1$ and $\epsilon_{max} = 15$ (11.76 dB).
The data power control is assumed to be on-off and only 
the analytical results (obtained by maximizing \eqref{eq:thpt}) 
are shown. (Note that the channel state pdf
is still given by Theorem \ref{thm}.)
At low SNRs the average training power and associated
overhead are small, so that taking the overhead into
account does not significantly affect the rate.
At high SNRs (around 10 dB)
the training overhead reduces the achievable rate by about $15\%$
with both the free and vertical boundaries.
The percentage improvement provided by the free boundary 
relative to the vertical boundary remains approximately the same.

A final remark is that when symbol overhead is taken into account,
the throughput associated with the vertical switching policy is
no longer the same as that associated with constant power control.
That is because the switching policy
requires on average $\epsilon_{avg}/\emax < 1$ channel uses 
for training every coherence block, whereas constant power control 
requires one channel use for training every coherence block. 
Of course, this savings in overhead for the switching policy
comes at the cost of feedback.

\section{Conclusions}
\label{conc}

We have studied achievable rates for
a correlated Rayleigh fading channel, where 
both the data and pilot power are adapted based 
on estimated channel gain.
In low SNR and fast fading scenarios the pilot power constitutes
a substantial fraction of the total power budget, so that
pilot power adaptation can provide
a substantial gain in achievable rates. 
By taking a diffusion limit, corresponding to low SNRs
(or wideband channel) and high correlation between consecutive 
channel realizations, several insights were obtained
about the optimal pilot power control policy.
Namely, it was shown that a policy that switches between
zero and peak training power is optimal, and that
the training power should be reduced when
the channel is bad and increased when the channel becomes good.
The optimal policy in the diffusion limit was also explicitly
characterized, and shown to provide a significant increase
in achievable rate for low SNRs and fast fading.

For the discrete-time system of interest the switching policy
is equivalent to maintaining constant pilot symbol power,
but inserting pilot symbols less frequently when 
the channel estimate is weak (and vice versa). 
When combined with on-off data power control,
this requires finite feedback, and achieves essentially 
the same performance with
the optimal (water-filling) data power control.
Of course, the CSI feedback required for optimal 
data and pilot power control can be substantially reduced
by exploiting the correlation between successive coherence blocks.

Several modeling assumptions have been made, which could 
be relaxed in future work. For example,
we have assumed that the receiver knows the statistical model
of the channel. In practice, the receiver may assume (or estimate) a
model, such as \eqref{eq:gm}, which is mismatched to
the actual channel statistics. An issue then is how sensitive
this overall performance is to this mismatch.
Also, the first-order Rayleigh fading model might be replaced
with other fading models
(e.g., Ricean, Nakagami, and higher-order autoregressive models).

Additional issues may arise when considering other channel models.
For example, here we have imposed
a power constraint, which is averaged over many coherence blocks. 
The results can therefore be directly applied to parallel
fading channels where the total power constraint is split
among the channels. However, for a frequency-selective channel
the total power summed over parallel channels
might instead be constrained per coherence block.
Other extensions and applications of diffusion models
to Multi-Input Multi-Output (MIMO) and multiuser channels
remain to be explored.





\appendices
\section{Continuous-time limit of discrete-time processes \eqref{eq:gm},
 \eqref{eq:kalest} and \eqref{eq:kalvar}}
\label{ap:lmt}

Substituting $r = 1- \rho\, \delta t$ into \eqref{eq:gm} 
and ignoring terms with higher power of $\delta t$, we obtain
\begin{equation}
h_{i+1} - h_{i} = - \rho \, \delta t \,h_i + \sqrt{2 \rho \,\delta t}\, w_i
\end{equation}
In the diffusion limit the noise
$\sqrt{\delta t}\,w_i$ can be modeled as $dB(t)$, 
where $B(t)$ is a standard complex Brownian Motion.
Hence as $\delta t \to 0$,
the preceding equation becomes \eqref{eq:ou}.


Substituting $r = 1-\rho\,\delta t$ in \eqref{eq:condvar} gives
\begin{equation}\label{eq:condvarcont}
\theta_{i+1|i} = (1- 2\rho\, \delta t )\,\theta_i + 2\rho\, \delta t\,{\sigma_h}^2.
\end{equation} 
Replacing $\epsilon_i$ by $\epsilon_i/N$ in \eqref{eq:kalvar}
and substituting $\delta t$ for $M/N$ gives
\begin{equation} \label{eq:condvarcont1}
\theta_{i+1} = (1 - \frac{\theta_{i+1}\epsilon_i \, \delta t}{{\sigma_z}^2})\, \theta_{i+1|i}.
\end{equation} 
Combining  \eqref{eq:condvarcont} and \eqref{eq:condvarcont1},
ignoring the $(\delta t)^2$ term, gives
\begin{equation}
\theta_{i+1} - \theta_{i} =  2\rho \,\delta t\,({\sigma_h}^2 -\theta_i )-  \frac{\theta_{i+1}\theta_{i}\epsilon_i \, \delta t}{{\sigma_z}^2} ,
\end{equation}   
which becomes \eqref{eq:errdyn} as $\delta t \to 0$.

Substituting for $r$ and replacing $\epsilon_i$ by 
$\epsilon_i/N$, \eqref{eq:kalest} can be re-written as,
\begin{equation}
\hat{h}_{i+1} - \hat{h}_{i} = - \rho\,\hat{h}_i\,\delta t + \frac{\theta_{i+1}}{{\sigma_z}^2}\,\left[ \epsilon_i \, 
\delta t (h_{i+1}-\hat{h}_i - \rho \, \delta t\,\hat{h}_i) + \sqrt{\epsilon_i\, \delta t \, {\sigma_z}^2}\,\, n_{i+1} \right]
\end{equation}
where, $n_{i+1}$ is a zero mean unit variance CSCG random variable
independent of $w_i$. 
The term $\frac{\theta_{i+1}}{{\sigma_z}^2}\, \epsilon_i \, 
\delta t (h_{i+1}-\hat{h}_i)$ is a CSCG random variable with mean zero 
and variance $\frac{\theta_{i+1}^2}{{\sigma_z}^4}\, \epsilon_i^2 \, 
E[|h_{i+1}-\hat{h}_i|^2] (\delta t)^2 $ and hence can be ignored.
Modeling $\sqrt{\delta t}\,n_{i+1}$ as $d\overline{B}(t)$, 
where $\overline{B}(t)$ is 
a standard complex Brownian motion independent of $B(t)$
gives \eqref{eq:chan} as $\delta t \to 0$.

\section{Proof of Lemma \ref{lm:cont}}
\label{ap:cont}
Defining the $3 \times 1$ state vector as 
$\mathbf{G}(t) = [\hat{h}_r(t), \,\,\, \hat{h}_j(t), \,\,\, \theta(t)]^\dag$,
\eqref{eq:chan} and \eqref{eq:errdyn} can be re-written as,
\begin{equation}\label{eq:VecSde}
d\mathbf{G}(t)  =  \mathbf{D}[\hat{h}_r(t), \hat{h}_j(t), \theta(t)]\, dt 
+ \mathbf{V}[\hat{h}_r(t), \hat{h}_j(t), \theta(t)]
\mathbf{\bar{B}}(t)
\end{equation}
where the drift and variance are given by
\begin{eqnarray}
\mathbf{D}(\hat{h}_r, \hat{h}_j, \theta) &=& \left[- \rho \,\hat{h}_r, \,\,\,- \rho \,\hat{h}_j,\,\,\, - 2\rho \,\theta  -
\frac{\epsilon  {\theta}^2 }{{\sigma_z}^2} + 2\rho\,{\sigma_h}^2\right]^\dag \\
\mathbf{V}(\hat{h}_r, \hat{h}_j, \theta) &=& \text{diag} \left[\theta \sqrt{\frac{\e}{2\sz}}, \,\, \theta \sqrt{\frac{\e}{2\sz}}, \,\, 0\right]
\end{eqnarray}
respectively, the dependence on time is dropped for notational convenience,
and the three entries of the vector 
$\mathbf{\bar{B}}(t)$ are independent, real-valued, 
standard Brownian motions.

From \cite[Theorem 5.2.1]{Oksend03}, given \eqref{eq:conlin} and 
\eqref{eq:conlip}, the solution to \eqref{eq:VecSde}
exists and is continuous in $t$ provided that
the following two conditions are satisfied:
\begin{equation}\label{eq:conlin}
|\mathbf{D}(\hat{h}_r, \hat{h}_j, \theta)| + |\mathbf{V}(\hat{h}_r, \hat{h}_j, \theta)| 
\le  C_1 (1 + \sqrt{\hat{h}_r^2 +\hat{h}_j^2 + \theta^2})
\end{equation}
\begin{multline}\label{eq:conlip}
|\mathbf{D}(\hat{h}_{r1}, \hat{h}_{j1}, \theta_1)- \mathbf{D}(\hat{h}_{r2}, \hat{h}_{j2}, \theta_2)| 
+ |\mathbf{V}(\hat{h}_{r1}, \hat{h}_{j1}, \theta_1)- \mathbf{V}(\hat{h}_{r2}, \hat{h}_{j2}, \theta_2)| \\
\le C_2 \sqrt{(\hat{h}_{r1}-\hat{h}_{r2})^2+ (\hat{h}_{j1}-\hat{h}_{j2})^2+(\theta_1-\theta_2)^2},
\end{multline}
where for any matrix $\mathbf{M}$ with $(k,l)$ entry $\mathbf{M}_{kl}$,
$|\mathbf{M}| = \sqrt{\sum \mathbf{M}_{kl}^2}$, 
and $C_1, C_2$ are constants. 
Condition \eqref{eq:conlin} is called the linear dominance property 
and \eqref{eq:conlip} is called the Lipschitz property.

Given $0\le \e \le \emax$, we have
\begin{eqnarray}
|\mathbf{D}| + |\mathbf{V}| &=& \sqrt{\rho^2 \hat{h}_r^2 + \rho^2 \hat{h}_j^2 +  \left(- 2\rho \,\theta  -
\frac{\epsilon  {\theta}^2 }{{\sigma_z}^2} + 2\rho\,{\sigma_h}^2\right)^2} +\sqrt{\frac{\e\theta^2}{\sz}}\\
&\le& \sqrt{\left(2\rho+\frac{\emax\sh}{\sz}\right)^2\left[\hat{h}_r^2+ \hat{h}_j^2+\theta^2\right]+ 4\rho^2 \sigma_h^4}
+ \sqrt{\frac{\emax\theta^2}{\sz}}\\
&\le& \left(2\rho+\frac{\emax\sh}{\sz}\right) \sqrt{\hat{h}_r^2+ \hat{h}_j^2+\theta^2} + \sqrt{4\rho^2 \sigma_h^4}+ \sqrt{\frac{\emax\theta^2}{\sz}}.
\end{eqnarray}
so that \eqref{eq:conlin} is satisfied. 
Similarly for $\theta_1 \ge \theta_2$ we have
\begin{multline}
|\mathbf{D}(\hat{h}_{r1}, \hat{h}_{j1}, \theta_1)- \mathbf{D}(\hat{h}_{r2}, \hat{h}_{j2}, \theta_2)| 
+ |\mathbf{V}(\hat{h}_{r1}, \hat{h}_{j1}, \theta_1)- \mathbf{V}(\hat{h}_{r2}, \hat{h}_{j2}, \theta_2)| \\
\le \sqrt{\rho^2(\hat{h}_{r1} - \hat{h}_{r2})^2 + \rho^2(\hat{h}_{j1} - \hat{h}_{j2})^2
+ \left[2\rho+\frac{2\emax\sh}{\sz}\right]^2(\theta_1-\theta_2)^2} + \sqrt{\frac{\emax}{\sz}(\theta_1-\theta_2)^2}.
\end{multline}
so that is \eqref{eq:conlip} satisfied.
Since the solution to \eqref{eq:estevol1} and \eqref{eq:errdyn}, 
$S(t) = (\mhat(t), \theta(t))$,
is a continuous function of $\mathbf{G}(t)$, 
it must also be continuous in $t$.

\section{Alternative Derivation of Continuous-Time Bellman Equation \eqref{eq:bellmid}}
\label{ap:alt}

We first rewrite the discrete-time Bellman equation \eqref{eq:bell1} as 
\begin{equation}\label{eq:bell1m}
 C = \max_{(P, {\epsilon})}\,\left\{ 
R(P, \hat{\mu}, \theta) - \lambda \,({\epsilon} 
 + P ) + E_{\epsilon, (\hat{\mu}, \theta)} [V] - V(\hat{\mu}, \theta)  \right\}, 
\end{equation}
where
\begin{equation}\label{eq:vdiff}
E_{\epsilon, (\hat{\mu}, \theta)} [V] - V(\hat{\mu}, \theta) =  
\int_0^{\infty}\, \left[V(u, \theta_{i+1}) - V(\hat{\mu}, \theta)  \right]\, 
f_{\hat{\mu}_{i+1} | S_i}(u) du .
\end{equation}

Assuming that $V(\hat{\mu}, \theta)$ is a continuous and smooth function, 
we can expand $V$ around $(\hat{\mu}, \theta)$ 
via the Taylor series
\begin{multline}\label{eq:srs}
V(u, \theta_{i+1}) - V(\hat{\mu}, \theta) = 
\frac{\partial V}{\partial \hat{\mu}} (u - \hat{\mu}) + 
\frac{\partial V}{\partial \theta} (\theta_{i+1} - \theta)\\
+  \frac{1}{2} \left[  \frac{\partial^2 V}{\partial \hat{\mu}^2} 
(u - \hat{\mu})^2 + 2 \frac {\partial^2 V}{\partial \hat{\mu} 
\partial \theta} (u - \hat{\mu})(\theta_{i+1} - \theta) 
+ \frac{\partial^2 V}{\partial \theta^2} 
(\theta_{i+1} - \theta)^2\right]\\
+ \textrm{higher-order terms}
\end{multline}
where all the derivatives are computed at $(\hat{\mu}, \theta)$. 
As stated in Sec. \ref{discrete},
$\hat{\mu}_{i+1}$ conditioned on $S_i$ is Ricean, so that
\begin{equation}\label{eq:rice}
f_{\mhat_{i+1} | S_i}(u) = 
\frac{1}{{\sigma_o}^2} e^{- (r^2\,\hat{\mu} + u)/{\sigma_o}^2} \, 
I_0 \bigg(\frac{2\sqrt{r^2 \hat{\mu} u }}{{\sigma_o}^2}  \bigg), 
\qquad  u \ge 0
\end{equation}
where $I_0(\cdot)$ is the zeroth-order modified Bessel function 
of the first kind and
\begin{equation}\label{eq:evar}
{\sigma_o}^2 = \frac{\epsilon M \theta_{i+1|i}}{{\sigma_z}^2} \theta_{i+1}
\end{equation} 
where $\theta_{i+1}$ and $\theta_{i+1|i}$ are given by \eqref{eq:kalvar} and 
\eqref{eq:condvar}, respectively, with $\theta_i$ replaced by $\theta$. 
The first two moments are \cite[Ch. 2]{Proaki08},
\begin{eqnarray}
\label{eq:mom1}
E[\hat{\mu}_{i+1} | (\hat{\mu},\theta) ] &=& r^2\,\hat{\mu} + {\sigma_o}^2\\
E[\hat{\mu}_{i+1}^2 | (\hat{\mu},\theta)] &=& 2 
{\sigma_o}^4\,\left[1 + 2 \bigg( \frac{r^2 \hat{\mu}}{{\sigma_o}^2} \bigg) + 
\frac{1}{2} \bigg( \frac{r^2 \hat{\mu}}{{\sigma_o}^2} \bigg)^2 \right]
\label{eq:mom2}
\end{eqnarray}

Next we take the diffusion limit. 
Substituting $r = 1 - \rho \,\delta t$ in \eqref{eq:evar}
and replacing $\epsilon$ by $\epsilon/N$ gives
\begin{equation}
{\sigma_o}^2 = \frac{\theta^2 \epsilon}{{\sigma_z}^2}\, (\delta t)
+ O(\delta t^2)
\end{equation}
Making these substitutions in \eqref{eq:mom1}-\eqref{eq:mom2} gives
\begin{eqnarray}
\label{eq:ltest}
E[\mhat_{i+1} - \hat{\mu} | (\hat{\mu},\theta)] &=& 
\left[ - 2 \rho \hat{\mu} + {\theta}^2  \frac{\epsilon}{{\sigma_z}^2}\right] \delta t + O(\delta t^2)\\
E[(\mhat_{i+1} - \hat{\mu})^2 | (\hat{\mu},\theta)] &=& 
\left[ 2 \hat{\mu} {\theta}^2  \frac{\epsilon}{{\sigma_z}^2}\right] \delta t
+ O(\delta t^2 )
\end{eqnarray}
It is easily shown that the higher-order moments
$E[(\hat{\mu}_{i+1} - \hat{\mu})^n | (\hat{\mu},\theta)] \leq O(\delta t^2)$
for $n \geq 2$, hence we can ignore the higher-order 
terms in \eqref{eq:srs}.

Substituting $\theta_i = \theta$ and taking the diffusion limit,  
\eqref{eq:kalvar} can be re-written as 
\begin{equation}\label{eq:ltvar}
\theta_{i+1} - \theta =  
\left[ - 2 \rho \theta + 2 \rho {\sigma_h}^2 - 
{\theta}^2  \frac{\epsilon }{{\sigma_z}^2} \right] \delta t
\end{equation}
Substituting \eqref{eq:srs} into \eqref{eq:vdiff} and
combining with \eqref{eq:ltest}-\eqref{eq:ltvar} gives
\begin{multline}\label{eq:vdifffinal}
E_{\epsilon, (\hat{\mu}, \theta)} [V] - V(\hat{\mu}, \theta) =   
\frac{\partial V}{\partial \hat{\mu}} 
\left[ - 2 \rho \hat{\mu} + {\theta}^2  
\frac{\epsilon}{{\sigma_z}^2} \right] \delta t \\
+ \frac{\partial V}{\partial \theta} 
\left[ - 2 \rho \theta + 2 \rho {\sigma_h}^2 - 
{\theta}^2  \frac{\epsilon}{{\sigma_z}^2} \right] \delta t 
+ \frac{\partial^2 V}{\partial \hat{\mu}^2} \left[ {\theta}^2  
\frac{\epsilon}{{\sigma_z}^2}\,\hat{\mu} \right] \delta t
\end{multline}
Lastly, $C$, $R$, $\e$ and $P$ can be multiplied by 
$\delta t$ without changing the original optimization problem. 
Applying this scaling in \eqref{eq:bell1m}, substituting 
\eqref{eq:vdifffinal} into \eqref{eq:bell1m},
and multiplying the entire equation by $1/\delta t$ 
and letting $\delta t \to 0$
gives \eqref{eq:bellmid}.

\section{Proof of Proposition \ref{lm:strip}}
\label{ap:strip}
Define the \emph{distance} from the free 
boundary at time $t$ as $\kappa(t) = \theta(t) - \theta_{\e}(\mhat(t))$. 
Irrespective of the 
initial state, due to the drift term in \eqref{eq:errdyn} and 
time continuity of the state process (Lemma \ref{lm:cont}), with probability one
there exists a finite time instant such that the state lies on the 
free boundary. Without loss of generality, rename that instance as
$t =0$ so that $\kappa(0) = 0$. For any $\eta >0$
let $t_1 = \inf \{t > 0: \kappa(t) \le - \eta\}$.
By continuity of the state process, $\kappa(t_1) = -\eta$ 
and there exists a
$t_0 = \sup\{ t \in (0, t_1): \kappa(t_0) = -\eta/2 \}$.

If $\kappa(t)<0$, bang-bang control implies that $\e(t) = 0$ and the
dynamical equations \eqref{eq:errdyn} and \eqref{eq:estevol1}
simplify to $d\theta = 2\rho(\sh-\theta) dt$ and 
$d\mhat = -2\rho \mhat~ dt$. Then
\begin{equation}
d\kappa = 2\rho\left[(\sh-\theta) - \mhat \frac{d\theta_{\e}(\mhat)}{d\mhat}\right]dt,
\end{equation}
and since $\kappa(t)<0$ implies $\theta<\theta_\e(\mhat)$, we have
$d\kappa > 2\rho\left[(\sh-\theta_{\e}) - 
\mhat \frac{d\theta_{\e}}{d\mhat}\right]dt$.
Thus given condition \eqref{eq:dervcon},  we have $d\kappa > 0$
whenever $\kappa(t) <0$. However, this contradicts the fact that 
$\int_{t_0}^{t_1} d\kappa = \kappa(t_1) -\kappa(t_0) = -\eta/2$.
Therefore we cannot have a $t_1 < \infty$, which implies that
$\lim_{t \to \infty}\Pr\{ \theta(t) \le \theta_\e(\mhat(t)) - \eta\} = 0$ 
for any $\eta >0$.

Next we show that 
$\lim_{t\to\infty}\Pr\{\theta(t) \ge \theta_\e(\mhat(t)) + \eta \} =0$.
For any continuous and twice differentiable 
function $W(\theta, \mhat)$ we must have \cite[Ch. 7]{Oksend03}
\begin{equation}\label{eq:ssp-cond}
E\left\{ A_{\epsilon}[W(\theta, \mhat)] \right\} = 0,
\end{equation}
where the expectation is over the steady-state
distribution of the state $(\theta, \mhat)$. 
The generator $A_{\epsilon}[\cdot]$ is defined as in \eqref{eq:gen},
except that the function $V(\cdot)$ is replaced by
$W(\cdot)$.
We choose $W(\cdot)$ to be a function of
$\theta$ only, i.e., $W(\theta, \mhat) = W(\theta)$, so that
\begin{equation}
A_{\epsilon}[W] = A_{\epsilon}[W(\theta)] = W' ( \theta) \left[2 \rho (\sh-\theta) -
  \frac{\epsilon \theta^2}{\sz}\right]
\end{equation}
where $W' (\theta)$ denotes the derivative of 
$W(\theta)$ with respect to $\theta$.
Let $p(u)$ denote the steady-state probability of training 
given that the channel estimate is $u$, as in \eqref{eq:ProbTrain}.
Rewriting \eqref{eq:ssp-cond} as
\begin{equation}
E_{\mhat}E_{\theta|\mhat}\left[A_{\epsilon}[W(\theta)] \right] = 0 ,
\end{equation}
and evaluating the inner conditional expectation
using the preceding result that 
$\Pr\{\theta \le \theta_\e(\mhat) - \eta | \mhat\} =0$
for any $\eta>0$ gives
\begin{multline}\label{eq:exp-temp}
E_{\mhat}\left\{ (1-p(\mhat)) W' (\theta_\e(\mhat)) 
\left[2 \rho (\sh-\theta_\e(\mhat))\right] \right\} \\
+ E_{\mhat} \left\{ p(\mhat) 
E_{\theta|\mhat;\theta\ge\theta_\e(\mhat)}
\left[W'(\theta)\left(2 \rho (\sh-\theta) -
\frac{\epsilon_{max} \theta^2}{\sz}\right)\right]\right\} = 0
\end{multline}
where $E_{\theta|\mhat;\theta \ge \theta_\e(\mhat)}[\cdot]$ 
denotes the expectation over $\theta$ given that the 
estimate is $\mhat$ and $\theta \ge \theta_\e(\mhat)$.  
Choosing $W(\theta)$ such that 
$W'(\theta) = 1/\left[2 \rho (\sh-\theta) - 
\frac{\epsilon_{max} \theta^2}{\sz}\right]$
and substituting in \eqref{eq:exp-temp} gives
\begin{equation}
E_{\mhat}\left[(1-p(\mhat)) 
\frac{2 \rho (\sh-\theta_\e(\mhat))}{2 \rho (\sh-\theta_\e(\mhat)) 
- \frac{\epsilon_{max} \theta_\e(\mhat)^2}{\sz}} \right]
+ E_{\mhat} [p(\mhat)] = 0 ,
\end{equation}
which implies
\begin{equation}\label{eq:exp-temp1}
E_{\mhat}\left[(1-p(\mhat)) 2 \rho (\sh-\theta_\e(\mhat)) \right]
= E_{\mhat} \left[p(\mhat)\left[\frac{\epsilon_{max} 
\theta_\e(\mhat)^2}{\sz} - 2 \rho (\sh-\theta_\e(\mhat)) \right]\right] .
\end{equation}
Next choose $W(\theta) = \theta$ so that $W'(\theta) = 1$. 
For this choice \eqref{eq:exp-temp} gives
\begin{equation}\label{eq:exp-temp2}
E_{\mhat}\left[(1-p(\mhat)) \left[2 \rho (\sh-\theta_\e(\mhat))\right] \right] 
= E_{\mhat} \left[p(\mhat) E_{\theta|\mhat;\theta \ge \theta_\e(\mhat)}
\left(\frac{\epsilon_{max} \theta^2}{\sz} -2 \rho (\sh-\theta)\right)\right].
\end{equation}

We now argue by contradiction that for any $\eta>0$,
$\Pr\{\theta \ge \theta_\e(\mhat) + \eta | \mhat\} =0$ 
almost everywhere (a.e.) in the set $\mathcal{M}= \{\mhat : \mhat > 0\}$.
If this were not the case, then we must
have $p(\mhat) >0$ 
over a subset in $\mathcal{M}$ with positive measure.
Since $\frac{\epsilon_{max} \theta^2}{\sz} - 
2 \rho (\sh-\theta)$ is a positive increasing 
function of $\theta$, \eqref{eq:exp-temp2} implies
\begin{equation}\label{eq:exp-temp3}
E_{\mhat}\left[(1-p(\mhat)) 2 \rho (\sh-\theta_\e(\mhat)) \right]
> E_{\mhat} \left[ p(\mhat)\left(\frac{\epsilon_{max} 
\theta_\e(\mhat)^2}{\sz} - 2 \rho (\sh-\theta_\e(\mhat)) \right)\right]
\end{equation}
with strict inequality, 
which contradicts \eqref{eq:exp-temp1}. 
Hence this establishes the proposition for any $\eta>0$.



\section{Proof of Theorem \ref{SSP}}
\label{ap:ssp}
As in Appendix \ref{ap:strip}, we use the fact that
for any continuous and twice differentiable
function $W(\theta, \mhat)$, we have \cite[Ch. 7]{Oksend03}
\begin{equation}\label{eq:ssp-cond1}
E\left\{ A_{\epsilon}[W(\theta, \mhat)] \right\} = 0,
\end{equation}
where the expectation is over the 
steady-state distribution of $(\theta, \mhat)$ and
the generator $A_{\epsilon}[\cdot]$ is given by \eqref{eq:gen}
with $V(\cdot)$ replaced by $W(\cdot)$.
Choosing $W(\mhat, \theta)= W_1(\theta)$ in \eqref{eq:ssp-cond1}
to be a function of $\theta$ only
and applying Proposition \ref{lm:strip} gives
\begin{equation}
\label{eq:pt-con}
E_{\mhat}\left[W_1' ( \theta_\e(\mhat)) g(\mhat)\right] = 0  
\end{equation}
where
\begin{equation}
g(\mhat) = (1-p(\mhat)) 2 \rho (\sh-\theta_\e(\mhat)) +\\ 
p(\mhat) \left[2 \rho (\sh-\theta_\e(\mhat)) -
\frac{\epsilon_{max} \theta_\e(\mhat)^2}{\sz}\right]. 
\end{equation}
Next we observe that $g(\mhat) =0$ a.e. in the
set $\mathcal{M}= \{\mhat: \mhat \ge 0\}$.  
If this were not the case, then since $\theta_\e(x)$ is a 
one-to-one function, we could choose
$W_1(\theta)$ such that $W_1'(\theta_\e(\mhat)) = g(\mhat)$,
which would make the left-hand side of \eqref{eq:pt-con}
strictly positive.
Therefore setting $g(\mhat) =0$ gives the steady-state probability
of training given $\mhat$ shown in \eqref{eq:ProbTrain}.

We now solve for the steady-state pdf $f_{\mhat}(u)$. 
Choosing $W(\theta,\mhat) = W_2(\mhat)$, a continuous and
twice differentiable function of $\mhat$ only, and applying
the generator \eqref{eq:gen} gives
\begin{equation}
A_{\epsilon}[W_2] = - 2\rho \mhat W_2'(\mhat) +  
\left[W_2'(\mhat) + \mhat W_2''(\mhat)\right]
\,\frac{\epsilon \,\theta^2}{\sz}.  
\end{equation}
The necessary condition \eqref{eq:ssp-cond1} can now be written as
\begin{equation}\label{eq:ssp-cond-mhat}
\int_0^\infty \left[C(u)W_2'(u) + D(u) W_2''(u)\right]f_{\mhat}(u) du = 0,
\end{equation}
where
\begin{equation}
C(u) = -2\rho u + \frac{[\theta_\e(u)]^2\emax}{\sz}
p(u) \quad \textrm{and} \quad D(u) = u \frac{[\theta_\e(u)]^2\emax}{\sz}
p(u).
\end{equation}
We can further choose $W_2(u)$ to satisfy the following properties:
\begin{eqnarray}
W_2(0) = 0\\
\lim_{u \to \infty} C(u) f_{\mhat}(u) W_2(u) = 0\\
\lim_{u \to \infty} D(u) f_{\mhat}(u) W_2'(u) = 0\\
\lim_{u \to \infty} \frac{d[D(u) f_{\mhat}(u)]}{du} W_2(u) = 0
\end{eqnarray}
and using integration by parts we can re-write \eqref{eq:ssp-cond-mhat} as
\begin{equation}
\int_0^\infty W_2(u)
\left(\frac{d^2}{du^2}[D(u)f_{\mhat}(u)] - 
\frac{d}{du}[C(u)f_{\mhat}(u)]\right) du = 0.
\end{equation}
Since this condition must be satisfied for
any such $W_2( \cdot )$, we have
\begin{equation} \label{eq:aefm}
\frac{d^2}{du^2}[D(u)f_{\mhat}(u)] -
\frac{d}{du}[C(u)f_{\mhat}(u)] = 0,
\qquad \textrm{a.e.} \,\,\,u \ge 0.
\end{equation}
Substituting \eqref{eq:ProbTrain} into \eqref{eq:aefm} gives the
differential equation
\begin{equation}
\frac{d^2}{du^2}\bigg\{2\rho(\sh-\theta_\e(\mhat))f_{\mhat}(u)\bigg\} -
\frac{d}{du}\bigg\{[2\rho(\sh-\theta_\e(u)) - 2\rho u]f_{\mhat}(u)\bigg\} = 0, 
\quad \textrm{a.e. } \,\,\,u \ge 0
\end{equation}
which can be further simplified as
\begin{equation}
2\rho[\sh-\theta_\e(u)]u \frac{df_{\mhat}(u)}{du} +
2\rho u \left(1 - \frac{d \theta_\e(u)}{du}\right)f_{\mhat}(u) + K = 0,
\end{equation}
where $K$ is a constant. 
This is a first-order ordinary differential equation with solution
\begin{equation}\label{eq:fmcrude}
f_{\mhat}(u) = - K \exp[-I(u)]
\int_{0}^{u}\frac{\exp[I(t)]}{2\rho\left[\sh-\theta_\e(t)\right] t} dt
+ K_1 \exp[-I(u)],
\end{equation}
where
\begin{eqnarray}
I(u) &=& \int_0^{u} \frac{1- d\theta_\e(t)/dt}{\sh-\theta_\e(t)} dt\\
 &=& \int_0^{u} \frac{1}{\sh-\theta_\e(t)} dt +
 \log[\sh-\theta_\e(u)] - \log[\sh-\theta_\e(0)]
\end{eqnarray}
and $K_1$ is another constant, which needs to be determined. 
Since $f_{\mhat}(u)$ is a pdf, we must have 
$\lim_{u \to \infty} f_{\mhat}(u) = 0$, which implies $K =0$.
This is because the first integral in \eqref{eq:fmcrude} 
is unbounded, that is,
\begin{equation}
\int_{0}^{u}\frac{\exp[I(t)]}{2\rho(\sh-\theta_\e)t} dt \ge
\int_0^1\frac{1}{2\rho\sh t} dt = \infty.
\end{equation}
In addition we must have $\int_0^\infty f_{\mhat}(u) du = 1$,
which implies $K_1 = 1/(\sh-\theta_\e(0))$. 
Substituting these values into \eqref{eq:fmcrude} 
gives \eqref{eq:SSgeneral}.

\section{Derivation of \eqref{eq:fb1}-\eqref{eq:analytic_fbd}}
\label{prop-wf-fbd}
First we fix the free boundary $\theta_\e(\mhat)$ and optimize
the data power allocation. For any $\mhat \ge 0$ setting the derivative of
the objective function \eqref{eq:fbd_lag} with respect
to $P(\mhat)$ to zero gives the optimal power allocation 
$P^\star(\mhat) = N P_d(\mhat, \theta_\e(\mhat), \lambda)$.
Substituting this $P^\star(\mhat)$ into \eqref{eq:fbd_lag} 
and taking the derivative with respect to $\theta_\e(\mhat)$
gives the optimality condition
\begin{multline}\label{eq:derv-theta}
\left[\frac{\partial L_{\lambda}}{\partial P}
[P^\star(\mhat), \mhat, \theta_\e(\mhat)] 
\cdot \frac{\partial P^\star(\mhat)}{\partial \theta_\e(\mhat)} +
 \frac{\partial L_{\lambda}}{\partial \theta} 
[P^\star(\mhat), \mhat, \theta_\e(\mhat)] \right]f_{\mhat}(\mhat)
= \\ - L_{\lambda}[P^\star(\mhat), \mhat, \theta_\e(\mhat)] 
\frac{f_{\mhat}(\mhat)}{\sh - \theta_\e(\mhat)}
+ \frac{1}{[\sh-\theta_\e(\mhat)]^2}\int_{\mhat}^{\infty} 
L_{\lambda}[P^\star(v), v, \theta_\e(v)] f_{\mhat}(v) dv.
\end{multline}

Note that 
$P^\star(\mhat) =0$ for $\mhat \le \lambda \sz$ so that
$\frac{\partial P^\star(\mhat)}{\partial \theta_\e(\mhat)} = 0$. 
For $\mhat \ge \lambda \sz$ we have 
$\frac{\partial L_{\lambda}}{\partial P}
[P^\star(\mhat), \mhat, \theta_\e(\mhat)]  =0$.
Therefore \eqref{eq:derv-theta} reduces to 
\eqref{eq:analytic_fbd} with 
$\theta_{\epsilon}$ replaced by $\theta_f(\mhat)$. 
The additional constraint 
$\theta_\e(\mhat) \ge \theta^\star$ implies \eqref{eq:fb1}.

\section{Free Boundary Problem as a Quadratic Optimization}
\label{VarInq}


We first observe that \eqref{eq:bellfinal1}
can be written as the {\em variational inequality} \cite{Friedm82}
\begin{eqnarray}\label{eq:varinq}
C - J - a \geq 0 \nonumber\\
C - J - a - \epsilon_{max}(b-\lambda) \geq 0 \nonumber\\
(C - J - a )(C - J - a - \epsilon_{max}(b-\lambda)) = 0
\end{eqnarray}
A solution to \eqref{eq:varinq} is a solution to
\eqref{eq:bellfinal1} and vice versa.
Now consider the following optimization problem,
\[
\min\, w_0  \int v_1\,v_2 \, d\theta ~ du + 
\sum_{x \in X}\, w_x \int \left[ (\partial_x v_1)^2 + (\partial_x  v_2)^2\right]
d\theta du \nonumber\\
\]
\begin{eqnarray}\label{eq:contmin}
Subject\,\, to: \,\, C - J - a = v_1 \ge 0  \nonumber\\
v_1 - \epsilon_{max}(b - \lambda) = v_2 \ge 0
\end{eqnarray}
where $\partial_x v_i = \frac{\partial v_i}{\partial x} dx$ 
for $x \in X = \{\mhat, \theta\}$ and $i = 1,2 $. If $w_0 > 0$,
$w_{\theta} = 0$, and $w_{\hat{\mu}} = 0$, then the solution to
\eqref{eq:varinq} is a solution to \eqref{eq:contmin}. Also, a
solution to \eqref{eq:contmin} with zero objective value is a
solution to \eqref{eq:varinq}. The second term in the objective
function is included to regularize the numerical solution. The
effect of this term can be controlled by changing the weights
$w_{\theta}$ and $w_{\hat{\mu}}$. These weights affect both the
accuracy of the results and also the rate at which the non-linear
optimization algorithm converges. The training region
is where $v_1(\hat{\mu}, \theta) > 0 $.
Therefore the free boundary can be obtained by
solving \eqref{eq:contmin} numerically
given values for $\lambda$ and $\rho$.

\bibliographystyle{IEEEtran}

\begin{thebibliography}{10}
\providecommand{\url}[1]{#1}
\csname url@samestyle\endcsname
\providecommand{\newblock}{\relax}
\providecommand{\bibinfo}[2]{#2}
\providecommand{\BIBentrySTDinterwordspacing}{\spaceskip=0pt\relax}
\providecommand{\BIBentryALTinterwordstretchfactor}{4}
\providecommand{\BIBentryALTinterwordspacing}{\spaceskip=\fontdimen2\font plus
\BIBentryALTinterwordstretchfactor\fontdimen3\font minus
  \fontdimen4\font\relax}
\providecommand{\BIBforeignlanguage}[2]{{%
\expandafter\ifx\csname l@#1\endcsname\relax
\typeout{** WARNING: IEEEtran.bst: No hyphenation pattern has been}%
\typeout{** loaded for the language `#1'. Using the pattern for}%
\typeout{** the default language instead.}%
\else
\language=\csname l@#1\endcsname
\fi
#2}}
\providecommand{\BIBdecl}{\relax}
\BIBdecl

\bibitem{TseVis05}
D.~Tse and P.~Viswanath, \emph{Fundametals of Wireless Communication}.\hskip
  1em plus 0.5em minus 0.4em\relax Cambridge University Press, 2005.

\bibitem{GolVar97IT}
A.~J. Goldsmith and P.~P. Varaiya, ``Capacity of fading channels with channel
  side information,'' \emph{IEEE Trans. Inform. Theory.}, vol.~43, no.~6, pp.
  1986--1992, Nov. 1997.

\bibitem{CaiSha99IT}
G.~Caire and S.~Shamai, ``On the capacity of some channels with channel state
  information,'' \emph{Information Theory, IEEE Transactions on}, vol.~45,
  no.~6, pp. 2007--2019, Sep 1999.

\bibitem{Viswan99IT}
H.~Viswanathan, ``Capacity of markov channels with receiver csi and delayed
  feedback,'' \emph{Information Theory, IEEE Transactions on}, vol.~45, no.~2,
  pp. 761--771, Mar 1999.

\bibitem{LiuEli04ACC}
J.~Liu, N.~Elia, and S.~Tatikonda, ``Capacity-achieving feedback scheme for
  flat fading channels with channel state information,'' \emph{American Control
  Conference, 2004. Proceedings of the 2004}, vol.~4, pp. 3593--3598 vol.4, 30
  June-2 July 2004.

\bibitem{CaiTar99IT}
G.~Caire, G.~Taricco, and E.~Biglieri, ``Optimum power control over fading
  channels,'' \emph{Information Theory, IEEE Transactions on}, vol.~45, no.~5,
  pp. 1468--1489, Jul 1999.

\bibitem{KleGal01ISIT}
T.~Klein and R.~Gallager, ``Power control for the additive white gaussian noise
  channel under channel estimation errors,'' \emph{Information Theory, 2001.
  Proceedings. 2001 IEEE International Symposium on}, pp. 304--, 2001.

\bibitem{YooGol06IT}
T.~Yoo and A.~Goldsmith, ``Capacity and power allocation for fading mimo
  channels with channel estimation error,'' \emph{Information Theory, IEEE
  Transactions on}, vol.~52, no.~5, pp. 2203--2214, May 2006.

\bibitem{Schram98TCOM}
P.~Schramm, ``Analysis and optimization of pilot-channel-assisted bpsk for
  ds-cdma systems,'' \emph{Communications, IEEE Transactions on}, vol.~46,
  no.~9, pp. 1122--1124, Sep 1998.

\bibitem{Cavers91TVT}
J.~Cavers, ``An analysis of pilot symbol assisted modulation for rayleigh
  fading channels [mobile radio],'' \emph{Vehicular Technology, IEEE
  Transactions on}, vol.~40, no.~4, pp. 686--693, Nov 1991.

\bibitem{Medard00IT}
M.~Medard, ``The effect upon channel capacity in wireless communications of
  perfect and imperfect knowledge of the channel,'' \emph{Information Theory,
  IEEE Transactions on}, vol.~46, no.~3, pp. 933--946, May 2000.

\bibitem{OhnGia02TWC}
S.~Ohno and G.~Giannakis, ``Average-rate optimal psam transmissions over
  time-selective fading channels,'' \emph{Wireless Communications, IEEE
  Transactions on}, vol.~1, no.~4, pp. 712--720, Oct 2002.

\bibitem{HasHoc03IT}
B.~Hassibi and B.~Hochwald, ``How much training is needed in multiple-antenna
  wireless links?'' \emph{IEEE Trans. Inform. Theory.}, vol.~49, no.~4, pp.
  951--963, April 2003.

\bibitem{BdeAbo04ICC}
A.~Bdeir, I.~Abou-Faycal, and M.~Medard, ``Power allocation schemes for pilot
  symbol assisted modulation over rayleigh fading channels with no feedback,''
  in \emph{Conference Record of the International Conference on Communications
  (ICC)}, vol.~2, 20-24 June 2004, pp. 737--741 Vol.2.

\bibitem{DonTon04TSP}
M.~Dong, L.~Tong, and B.~Sadler, ``Optimal insertion of pilot symbols for
  transmissions over time-varying flat fading channels,'' \emph{IEEE
  Transactions on Signal Processing}, vol.~52, no.~5, pp. 1403--1418, May 2004.

\bibitem{WanMoa95TVT}
H.~S. Wang and N.~Moayeri, ``Finite-state markov channel-a useful model for
  radio communication channels,'' \emph{Vehicular Technology, IEEE Transactions
  on}, vol.~44, no.~1, pp. 163--171, Feb 1995.

\bibitem{WanCha96TVT}
H.~S. Wang and P.-C. Chang, ``On verifying the first-order markovian assumption
  for a rayleigh fading channel model,'' \emph{Vehicular Technology, IEEE
  Transactions on}, vol.~45, no.~2, pp. 353--357, May 1996.

\bibitem{TanBea00TCOM}
C.~Tan and N.~Beaulieu, ``On first-order markov modeling for the rayleigh
  fading channel,'' \emph{Communications, IEEE Transactions on}, vol.~48,
  no.~12, pp. 2032--2040, Dec 2000.

\bibitem{ZhaKas99TCOM}
Q.~Zhang and S.~Kassam, ``Finite-state markov model for rayleigh fading
  channels,'' \emph{Communications, IEEE Transactions on}, vol.~47, no.~11, pp.
  1688--1692, Nov 1999.

\bibitem{Bertse95}
D.~P. Bertsekas, \emph{Dynamic programming and optimal control}.\hskip 1em plus
  0.5em minus 0.4em\relax Athena Scientific, Vol. 1 and 2, 1995.

\bibitem{NegCio02IT}
R.~Negi and J.~Cioffi, ``Delay-constrained capacity with causal feedback,''
  \emph{IEEE Trans. Inform. Theory.}, vol.~48, no.~9, pp. 2478--2494, Sep 2002.

\bibitem{ZafMod05Allerton}
M.~Zafer and E.~Modiano, ``Continuos-time optimal control for delay constrained
  data transmission,'' in \emph{Allerton Conference on Communication, Control
  and Computing, Urbana, IL, USA, September 2005.}, 2005.

\bibitem{ChaDjo05IT}
C.~Charalambous, S.~Djouadi, and S.~Denic, ``Stochastic power control for
  wireless networks via sdes: probabilistic qos measures,'' \emph{Information
  Theory, IEEE Transactions on}, vol.~51, no.~12, pp. 4396--4401, Dec. 2005.

\bibitem{BerGal02IT}
R.~Berry and R.~Gallager, ``Communication over fading channels with delay
  constraints,'' \emph{IEEE Trans. Inform. Theory.}, vol.~48, no.~5, pp.
  1135--1149, May 2002.

\bibitem{Oksend03}
B.~Oksendal, \emph{Stochastic Differential Equations: An Introduction with
  Applications}.\hskip 1em plus 0.5em minus 0.4em\relax Springer-Verlag Berlin
  Heidelberg, Sixth Edition, 2003.

\bibitem{FenFie07TCOM}
T.~Feng, T.~Field, and S.~Haykin, ``Stochastic differential equation theory
  applied to wireless channels,'' \emph{IEEE Transactions on Communications},
  vol.~55, no.~8, pp. 1478--1483, Aug. 2007.

\bibitem{Zwilli98}
D.~Zwillinger, \emph{Handbook of differential equations}.\hskip 1em plus 0.5em
  minus 0.4em\relax Academic Press, Third Edition, 1998.

\bibitem{Friedm82}
A.~Friedman, \emph{Variational principles and free boundary problems}.\hskip
  1em plus 0.5em minus 0.4em\relax Wiley-Interscience, John Wiley and Sons,
  1982.

\bibitem{BroHwa97}
R.~G. Brown and P.~Hwang, \emph{Introduction to random signals and applied
  kalman filtering}.\hskip 1em plus 0.5em minus 0.4em\relax John Wiley and
  Sons, 1997.

\bibitem{AgaHon05Allerton}
M.~Agarwal and M.~L. Honig, ``Wideband fading channel capacity with training
  and partial feedback,'' in \emph{Allerton Conference on Communication,
  Control and Computing, Urbana, IL, USA, September 2005.}, 2005.

\bibitem{Whitt02}
W.~Whitt, \emph{Stochastic-Process Limits}.\hskip 1em plus 0.5em minus
  0.4em\relax Springer, 2002.

\bibitem{DynYus79}
E.~Dynkin and A.~A. Yushkevich, \emph{Controlled Markov Processes}.\hskip 1em
  plus 0.5em minus 0.4em\relax Springer Verlag, New York, 1979.

\bibitem{Bramso98QUE}
M.~Bramson, ``State space collapse with application to heavy traffic limits for
  multiclass queueing networks,'' \emph{Queueing Systems: Theory and
  Applications}, vol.~30, no.~2, pp. 89--148, 1998.

\bibitem{Rudin86}
W.~Rudin, \emph{Real and Complex Analysis}.\hskip 1em plus 0.5em minus
  0.4em\relax McGraw-Hill Series in Higher Mathematics, Third Edition, 1986.

\bibitem{SunHon08IT}
Y.~Sun and M.~L. Honig, ``Asymptotic capacity of multi-carrier transmission
  over a fading channel with feedback,'' \emph{IEEE Trans. Inform. Theory.},
  2008, to appear.

\bibitem{Proaki08}
J.~Proakis and M.~Salehi, \emph{Digital Communications}, 5th~ed.\hskip 1em plus
  0.5em minus 0.4em\relax McGraw-Hill, 2008.

\end{thebibliography}

\end{document}